\documentclass{aa}  
\usepackage{graphicx}
\usepackage[varg]{txfonts}
\usepackage{hyperref}
\usepackage{textgreek}

\newcommand{\ms}{\,\mathrm{M}_\odot}

\newcommand{\kms}{\,\mathrm{km/s}}
\newcommand{\days}{\,\mathrm{d}}
\newcommand{\vrot}{\varv_\mathrm{rot}}
\let\del\partial

\begin{document} 

\title{The spins of stripped B stars support\\magnetic internal angular momentum transport}

\author{C.~Schürmann\inst{1}\fnmsep\inst{2}\fnmsep\thanks{email: \texttt{chr-schuermann@uni-bonn.de}}
\and N.~Langer\inst{2}\fnmsep\inst{1}
\and X.~Xu\inst{1}\fnmsep\inst{2}
\and C.~Wang\inst{2}\fnmsep\inst{3}}

\institute{Max-Planck-Institut für Radioastronomie, Auf dem Hügel 69, 53121 Bonn, Germany
\and Argelander-Institut für Astronomie, Universität Bonn, Auf dem Hügel 71, 53121 Bonn, Germany
\and Max-Planck-Institut für Astrophysik, Karl-Schwarzschild-Str. 1, 85748 Garching, Germany}

\authorrunning{C. Schürmann et al.}

\date{Submitted 31\textsuperscript{st} May 2022 / Accepted 4\textsuperscript{th} August 2022}

\abstract{In order to predict the spins of stellar remnants we need to understand the evolution of the internal rotation of stars, and to identify at which stage the rotation of the contracting cores of evolved stars decouples from their expanding envelopes. The donor stars of mass transferring binaries lose almost their entire envelope and may thus offer a direct view on their core rotation. After the mass transfer event they contract and fade rapidly, although they are well observable when caught in the short-lived B-star phase. The B-type primary of the galactic binary system \object{LB-1}, which was originally suggested to contain a massive black hole, is nicely explained as a stripped star accompanied by a fainter Be star. The narrow absorption lines in the primary's spectrum signify extremely slow rotation, atypical of B-type main-sequence  stars. Here we investigate the evolution of mass donors in generic grids of detailed binary evolution models, where both stars include differential rotation, internal angular momentum transport, and spin-orbit coupling. Whereas the mass gainers are typically spun-up during the mass transfer, we find that the spins of the stripped donor models   depend sensitively on the employed mechanism for internal angular momentum transport. Purely hydrodynamic transport cannot explain the observed slow rotation, while models including magnetic angular momentum transport are able to reproduce the observed rotation of LB-1 and similar stars, independent of the initial rotation rate. In such models the spin of the white dwarfs that emerge at the end of the evolution is independent of the mass stripping. We find evidence that the mass transfer in LB-1 was moderately non-conservative.}


\maketitle

\section{Introduction}
It is well known that upper main-sequence stars are often rapid rotators. For a long time stellar rotation was considered  a second-order effect, but it turned out, most notably in O- and B-type stars, that it can strongly affect their evolution. While it may be at the core of observed phenomena such as gamma-ray bursts or luminous blue variables, numerical simulations still struggle to yield a convincing overall picture \citep[][and references therein]{2000ARA&A..38..143M, 2012ARA&A..50..107L}.
%other rotation stuff: mixing, CHE, wind enhancement, LBVs, SN types, WR
%other magnetic stuff: magnetars, spin down, transport, NS, GRB

Rotation may be faster in stars with companions \citep{2013ApJ...764..166D}. In a binary system, the two stars can interchange material and thus angular momentum. Furthermore, tidal forces, which grow with the stars' Roche-lobe filling-factors  \citep{1977A&A....57..383Z}, act on both of them. The internal rotational structure does not need to be uniform, but it may depend on the radial coordinate \citep{1992A&A...265..106S, 1992A&A...265..115Z}. Such a differential rotation is counteracted by turbulent viscosity \citep{heger2000} and magnetic fields \citep{spruit}, forcing a star close to rigid rotation \citep{2004A&A...422..225M}. Unfortunately these processes are hidden under the stellar surface. A direct look at the stellar core would be illuminating. In this context, stripped stars are of great interest as they provide the desired opportunity to look deep inside the stellar structure.

Recently, \citet{liu19} proposed that the galactic B-type binary LB-1 contains a $70\ms$ black hole (BH). They found an antiphase radial velocity variation of the thin absorption lines and the H\textalpha~emission lines in the composite spectrum. This, together with a typical mass for the early B-type star is the basis for the mass estimate, as the H\textalpha~emission was assumed to originate from the vicinity of the BH. This observation caused great interest in the community since such a massive BH was not expected to be able to form in a high metallicity environment \citep[e.g.][]{2002ApJ...567..532H, 2018MNRAS.481.1908K, 2020ApJ...905L..15B}. \citet{irrgang} examined the stellar absorption lines and found the B-type star to be a stripped helium star leading to a smaller BH mass or even just a neutron star. In contrast, \citet{simondiaz}  classified it as a slightly evolved main-sequence star with solar surface helium abundance. Both studies found an enrichment in CNO processed material. However a systematic mismatch between the observed line profiles and fits remained.

\citet{elbadry} and \citet{abdulmasih} state that the movement of the H\textalpha~emission line may not be real, but an effect of combining a stationary H\textalpha~line with a varying H\textalpha~absorption yielding an apparent movement of the combined line in antiphase with the absorption feature. According to \citet{elbadry} the unseen companion is a stellar mass BH and the H\textalpha~emission is caused by a circumbinary disc. However, the observations of \citet{liu20} out-ruled such a disc in the system.

A convincing overall picture was achieved by \citet{shenar}, who analysed the system by spectral disentangling. The authors were able to identify \emph{two} stellar components, one narrow-lined helium-enriched star and one rapidly rotating star, whose absorption lines are very broad and difficult to spot. This second star is also the source of the H\textalpha~emission as it is identified as a Be~star. The helium-rich star was found to be CNO-enriched, suggesting that it was the mass donor in a Roche-lobe overflow (RLO), in which the emission line star received mass and angular momentum, transforming it into a Be~star \citep{1989A&A...220L...1W, 2020ApJ...888L..12W, 2020A&A...633A..40L}. As the stripped B-type star (Bstr) has a temperature and surface gravity of a slightly evolved main-sequence star, it seems  likely that the RLO happened quite recently and that the stripped star is contracting towards a \textphi~Per-like OB subdwarf (sdOB) in orbit with a Be~star. \citet{lennon} tested this model and the original one \citep[B+BH,][]{liu19} with a Hubble UV-visual-IR spectrum and found that neither could reproduce all the observed properties.

The LB-1 system is apparently not the only one of its kind. \citet{bodensteiner} re-analysed the apparent (B+BH)+Be-system \object{HR~6819} first examined by \citet{rivinius} and found it to consist of a stripped B~star and a Be~star. \citet{elbadry2} came to the same result. \citet{eldridge}, \citet{bodensteiner}, and \citet{elbadry2} provide numerical evolutionary models for LB-1 and HR~6819. According to \citet{2022MNRAS.511L..24E}, \object{NGC~1850 BH1} found by \citet{2022MNRAS.511.2914S} is also a Bstr+Be system.

For this study, we adopt the Be+Bstr model for LB-1. We employ this model to examine the aforementioned processes influencing stellar rotation since the low mass of their envelope is very sensitive to core-envelope-coupling. As LB-1 (and HR~6819) is likely in a short-lived phase, a post-RLO contraction, the coupling could not have reached an equilibrium yet, which makes it a valuable target. Thus, the objectives of this study are first to identify a numerical model describing the stripped star of LB-1 and if possible the system as a whole, and second to use the predicted and observed rotational velocity to draw conclusions about the angular momentum transport mechanisms in the stellar interior.

In Sect.~\ref{sec:emp} of this paper we summarise the observed properties of LB-1 and place this kind of system into the context of binary stellar evolution. In Sect.~\ref{sec:method} we present our numerical method, and in Sect.~\ref{sec:results} present its results, the progenitor model of LB-1, and its rotational evolution. Finally, in Sect.~\ref{sec:discuss}, we discuss similar systems, earlier work, and the orbital evolution. We  draw our conclusions in Sect.~\ref{sec:concl}.

\section{Empirical properties of Be~stars with stripped companions}\label{sec:emp}

\begin{table}
    \centering
    \caption{Orbital and atmospheric properties of the stripped star in LB-1 according to \citet{shenar} and \citet{lennon} in the Be+Bstr scenario.}
    \label{tab:atmos}
    \begin{tabular}{lcc} \hline\hline
         & \citeauthor{shenar} & \citeauthor{lennon} \\ \hline
        $P_\mathrm{orb}\,\mathrm{/d}$ & $78.7999\pm0.0097$ & -- \\
        $q$ & $4.7\pm0.4$ & -- \\
        $M_\mathrm{tot} \sin^3 i \,\mathrm{/M_\odot}$ & $2.16\pm0.05$ & -- \\ \hline
        $T_\mathrm{eff}\,\mathrm{/K}$ & $12\,700\pm2\,000$ & $12\,500\pm100$ \\
        $\log g \,\mathrm{/\,cm/s^2}$ & $3.0\pm0.2$ & $3.0\pm0.2$ \\
        $\vrot \sin i \,\mathrm{/\,km/s}$ & $7\pm2$ & $7$ \\
        $n_\mathrm{He}/n_\mathrm{H}$ & $0.21$\tablefootmark{a} & $0.2$ \\ \hline
        $\log (L/M\,/\,\mathrm{L}_\odot/\mathrm{M}_\odot)$ & $2.81\pm0.34$ & $2.80\pm0.21$ \\
        $Y_\mathrm{surface}$ & $0.46$\tablefootmark{(a)} & $0.44\pm0.12$ \\
        $[\mathrm{N}/\mathrm{C}]$ & $>0$\tablefootmark{(b)} & $2.25\pm0.21$\tablefootmark{(c)} \\ \hline
    \end{tabular}
    \tablefoot{The last three lines were calculated by us. \tablefoottext{a}{Hawcroft \& Shenar (priv. com.) find $n_\mathrm{He}/n_\mathrm{H} = 0.31\pm0.05 \Leftrightarrow Y=0.55\pm0.04$.} \tablefoottext{b}{Hawcroft \& Shenar (priv. com.) find $[\mathrm{N}/\mathrm{C}] = 2.24\pm0.38$}. \tablefoottext{c}{Estimated from their Table~3.}}
\end{table}

We summarise in Table~\ref{tab:atmos} the empirical properties of the LB-1 system and the atmospheric properties of the stripped star according to the two studies discussing the Be+Bstr scenario. For our following analysis we adopt a surface temperature of 12\,500\,K and a surface gravity of 3.0. It is also clear that the stripped star's surface is enriched with helium and CNO-products.  \citet{shenar} and \citet{lennon} report $Y\approx0.45$, while Hawcroft \& Shenar (priv. com.) find $Y\approx0.55$. Two studies \citep{irrgang,simondiaz} assuming the B+BH model prefer $Y\approx0.65$ and $Y\approx0.29$ (i.e. solar), respectively. The scatter in the proposed helium abundances is quite large. We give a higher weight to the Be+Bstr models and use an interval of $Y\in[0.4, 0.6]$ for our study. For the relative abundance of nitrogen and carbon we adopt $2.0 < \mathrm{[N/C]} < 2.5$. Furthermore we assume a projected equatorial rotational velocity of the stripped star of $7\pm2\kms$.

A similar synopsis could be made for HR~6819 \citep{rivinius,bodensteiner,elbadry2}, but unfortunately only an upper limit for the rotation of the stripped star is known \citep[$\vrot\sin i < 20\kms$,][]{elbadry2}, which is not precise enough for   the analysis presented below. For NGC~1850 BH1 these measurements are not published.

LB-1, HR~6819, and NGC~1850 BH1 may be the first members of a new family of B-type stars, and Be~stars in particular. To date, only a small fraction of Be~stars are known to have a companion \citep{2020A&A...633A..40L} even though binary interaction is a proposed formation channel \citep[e.g.][]{1991A&A...241..419P, 2020ApJ...888L..12W}. On the other hand, all known companions are post-interaction objects and no Be~star with a main-sequence companion is know, although these stars should be easily detectable and a large number of B+B binaries are known \citep{2020A&A...641A..42B}.

Most known Be~star companions, besides neutron stars, are sdOB~stars, of which \citet{2018ApJ...853..156W} list 16 detections and candidates. The most prominent member of this family is \textphi~Per \citep[e.g.][]{1981PASP...93..297P,1998ApJ...493..440G,abel}. It is believed that these stars are the stripped cores of the mass donors that spun up the Be~stars to high rotation. Be+Bstr systems such as LB-1 and HR~6819 may evolve into  Be+sdOB systems \citep{shenar} as the donor star crosses the main sequence during its evolution from Roche-lobe filling to sdOB~star. It is also possible, but less likely because the timescale is  about ten times shorter  \citep{elbadry2}, that LB-1 and HR~6819 are in the evolutionary stage after core helium exhaustion of the sdOB star, when it expands again due to shell helium burning to become a helium giant. Recently, \citet{2022arXiv220105614E} proposed that  \object{HD~15124} is  a Be~star with a Roche-lobe filling companion, which will evolve first to a LB-1 like system and then to a sdOB+Be-system.

Although predicted by abundance \citep[about 70\%,][]{2001A&A...367..848R}, a white dwarf (WD) companion is proposed for only six Be~stars \citep{2021MNRAS.508..781K}. Moving up the mass ladder, many Be~stars with neutron stars are known as Be/X-ray binaries. Even one BH accompanying a Be~star is observed \citep{2014Natur.505..378C}. LB-1 and HR~6819 are likely the progenitors of the presumed Be+WD systems, as their stripped stars appear to have a mass of less than $1.5\ms$ \citep{shenar,lennon,bodensteiner}.

\section{Method}\label{sec:method}
The basis of our analysis is a large grid of detailed binary evolution models with the  Small Magellanic Cloud (SMC) metallicity calculated by \citet{wanginprep}, using MESA version 8845 \citep{mesa1,mesa2,mesa3}. Their initial zero age main sequence (ZAMS) equatorial velocity is set to 0.55 times the critical velocity, corresponding to the high velocity peak of the bimodal distribution of \citet{2013A&A...550A.109D}. The other initial binary properties were randomly drawn from empirical distributions (Monte Carlo method). Initial primary masses range from $3\ms$ to $100\ms$, mass ratios range from $1$ to $0.1$, and the initial orbital periods lie between initial contact and 3000\,d.

We set the stellar physical parameters as follows. The overshooting is assumed to be mass dependent \citep{2014A&A...570L..13C,2021A&A...648A.126M}. For $M<1.25\ms$ we use $\alpha_\mathrm{ov}=0$, for $1.25\ms<M<1.7\ms$ we set $\alpha_\mathrm{ov}=0.05$, and above $1.7\ms$ we follow \citet{2019A&A...625A.132S} by increasing $\alpha_\mathrm{ov}$ linearly from 0.1 at $1.7\ms$ to 0.3 at $20\ms$. Semiconvection is set to $\alpha_\mathrm{sc}=10$, as suggested by \citet{2019A&A...625A.132S}. In the case of mass transfer by stable RLO, we assume that the accretor gains mass until it reaches critical rotation \citep{2005A&A...435.1013P,mesa3}. Then the transferred material is expelled with the accretor's orbital angular momentum. This leads to an accretion efficiency of less than $5\%$ in systems where for the accretor \citep{2020A&A...638A..39L} tidal forces do not play a role (in general in RLOs after the donor has  left the main sequence). Whether the mass transfer is stable and all the material that cannot be accreted is ejected successfully or the binary undergoes a common envelope phase and merges is decided by an energy criterion. If the combined luminosity of the two stars is large enough to unbind excess material from the system, we assume the RLO avoids a common envelope \citep{pablo, 2020A&A...638A..39L}.

If no external factors such as stellar wind, accretion, and tides act on the model, its total spin angular momentum is conserved. MESA treats rotation by assigning an angular velocity to each mass shell of the stellar model \citep{mesa2}. The angular velocity $\omega$ of each shell can change by two means according to
\begin{equation} \label{eq:dodt}
    \frac{\del\omega}{\del t} = - \frac{\omega}{i}\frac{\del i}{\del t} + \frac1i \frac{\del}{\del m} \left( (4\pi r^2 \rho)^2 i \nu \frac{\del\omega}{\del m} \right).
\end{equation}
The first term is the change in specific moment of inertia $i$. The second, the change in specific angular momentum of a shell, is described by a diffusion ansatz \citep{heger2000} parametrised by a viscosity $\nu$, which serves as an effective description of all the physical processes involved in the coupling between the shells. The most important process is convection, which is assumed to impose rigid body rotation \citep{heger2000}. Semiconvection and thermohaline mixing are also included, as well as atomic viscosity. Rotation itself induces a set of instabilities leading to angular momentum transport: the dynamical shear instability, the Solberg-H\o iland instability, the secular shear instability, the Eddington-Sweet circulations, and the Goldstein-Schubert-Fricke instability \citep{heger2000}. An important contribution to the viscosity is assumed to be due to magnetic fields in the form of the Spruit--Tayler mechanism \citep{spruit, heger2005}. This magnetic viscosity depends on the $-4${th} power of the Brunt-Väisälä frequency, which means that gradients in entropy or mean molecular weight reduce the magnetic viscosity.

We use our grid of SMC models to infer possible initial masses, initial mass ratios, and initial orbital periods of LB-1, as described in Sect.~\ref{sec:res-pro}. However, the grid is  not dense enough to contain a model that closely matches the inferred properties. Therefore, we ran additional models with our favoured regime of the initial binary properties. For these we chose solar metallicity since LB-1 is a Milky Way binary. We found the differences between the SMC and Milky Way stripped stellar models to be small. In order to characterise the impact of the initial rotation, and of the magnetic angular momentum transport, we calculated solar metallicity models without magnetic angular momentum transport, and models rotating at 0.2 times their critical rotation velocity at ZAMS. Additionally, we calculated a single-star model of the primary using our fiducial physics.

\section{Results}\label{sec:results}
In this section we present our findings based on the set of SMC binary models and on the recalculated models at Milky Way metallicity. First we identify possible progenitor models of LB-1 in Sect.~\ref{sec:res-pro}, and continue a brief review of evolutionary stages with similar surface properties (Sect.~\ref{sec:res-other}). We then turn to the evolution of rotation and the transport of angular momentum in our solar metallicity model in Sect.~\ref{sec:res-rot}. We   discuss models with different initial rotations and without the Spruit--Tayler dynamo (Sect.~\ref{sec:res-alt}), and we compare our binary result to that of a single star (Sect.~\ref{sec:res-sing}). To round off we discuss predictions about systems similar to LB-1 (Sect.~\ref{sec:res-prog}).

\subsection{Progenitor models for LB-1}\label{sec:res-pro}
In order to identify the models in our model grid which most closely resemble LB-1 in the Be+Bstr scenario, we search for a contracting stripped stellar model in a binary system that underwent RLO without merging, with a mass donor with a temperature of about 12\,500\,K and a surface gravity of about 3.0 and which has not yet depleted central helium burning. The last criterion means that it is contracting towards the sdOB~phase.

\begin{figure}
    \includegraphics[width=1.1\hsize]{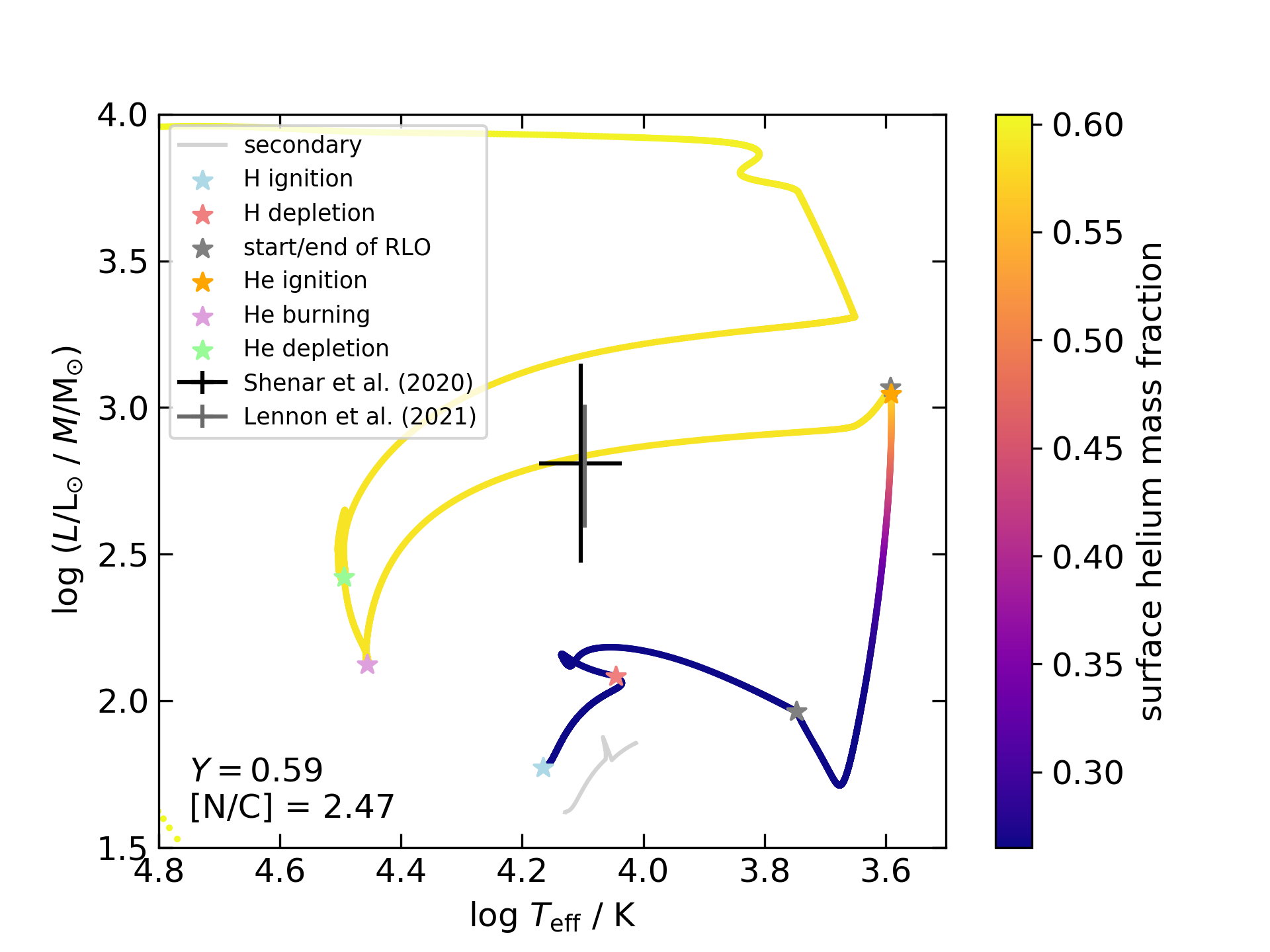}
    \caption{Spectroscopic Hertzsprung-Russell diagram of a possible LB-1 progenitor. The helium abundance of the donor is indicated by colour. The mass gainer, which does not leave the main sequence as the binary model ends when the primary becomes a WD, is shown in grey. The  ZAMS parameters are $M_1 = 4.0\ms$, $M_2 = 3.5\ms$, and $P_\mathrm{orb} = 16\days$. After the RLO the parameters were $M_1=0.7\ms$, $M_2=3.5\ms$, and $P_\mathrm{orb} = 223\days$. The surface abundances of the donor after RLO are shown in the lower left corner. The star symbols indicate Roche-lobe decoupling, helium ignition, middle of helium burning phase (sdOB observationally), and central helium depletion. The observations of \citet{shenar} and \citet{lennon} are shown in black and grey.}
    \label{fig:hrd}
\end{figure}

\begin{figure*}
    \includegraphics[width=0.5\hsize]{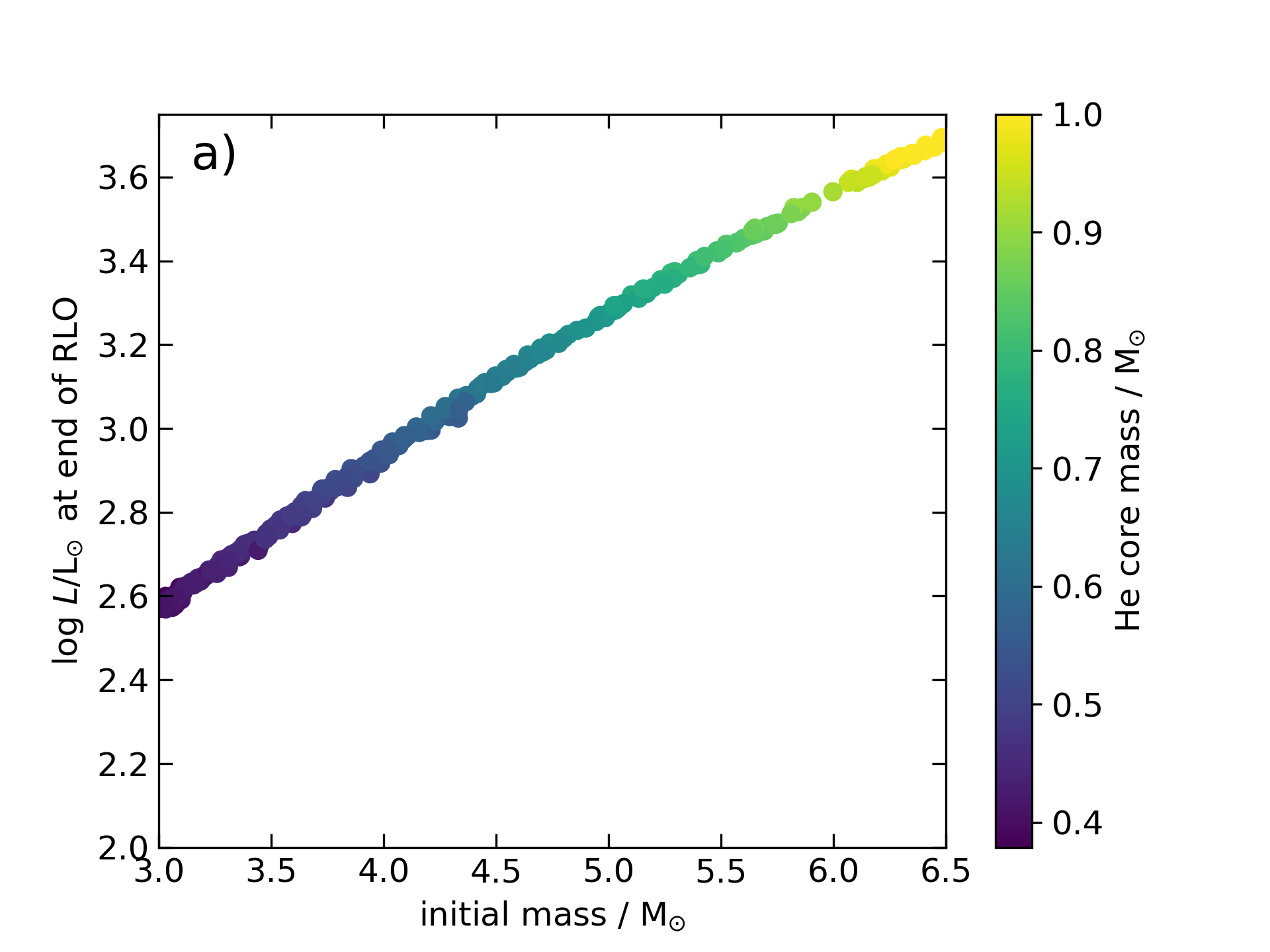}
    \includegraphics[width=0.5\hsize]{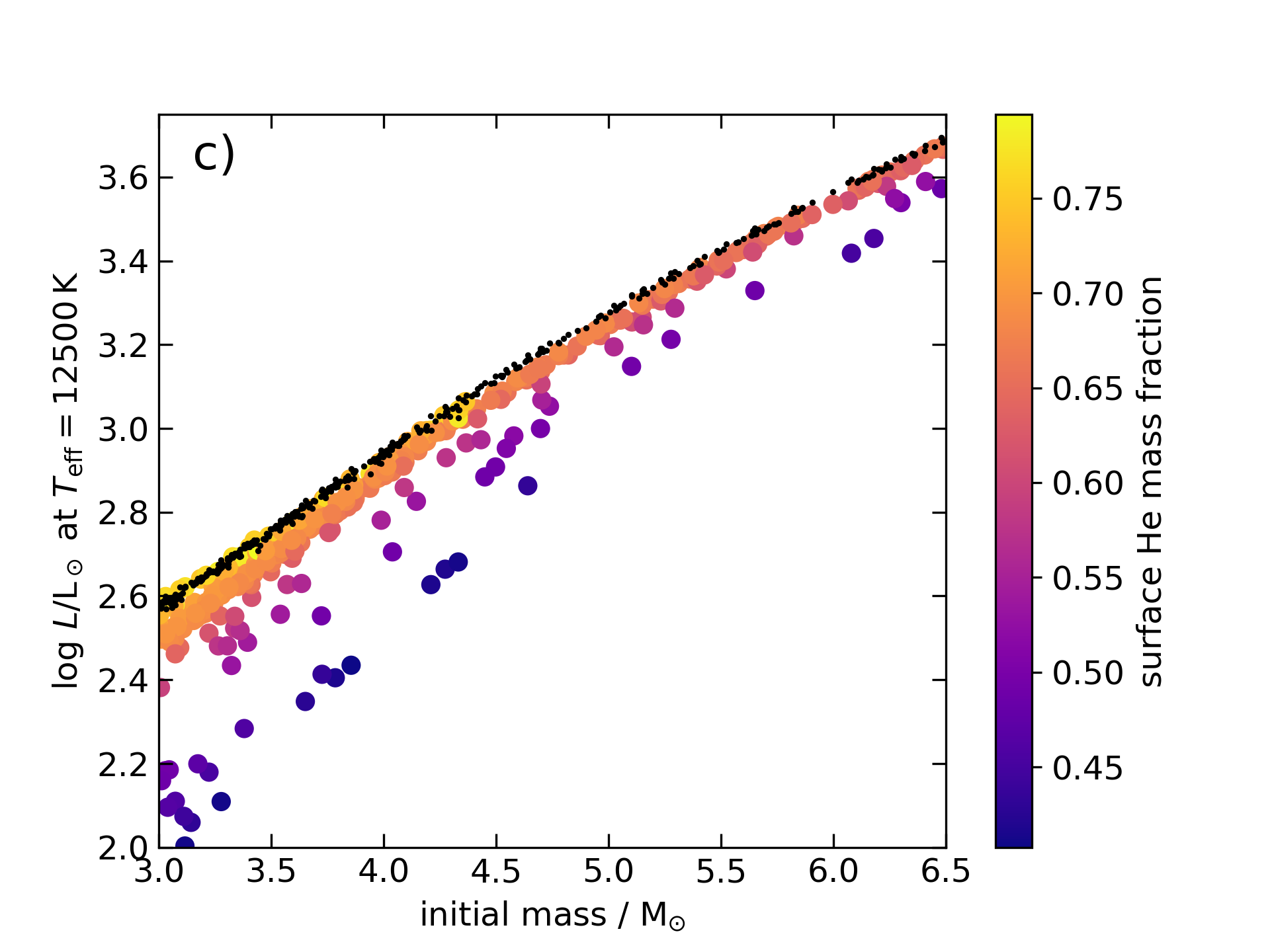}
    \includegraphics[width=0.5\hsize]{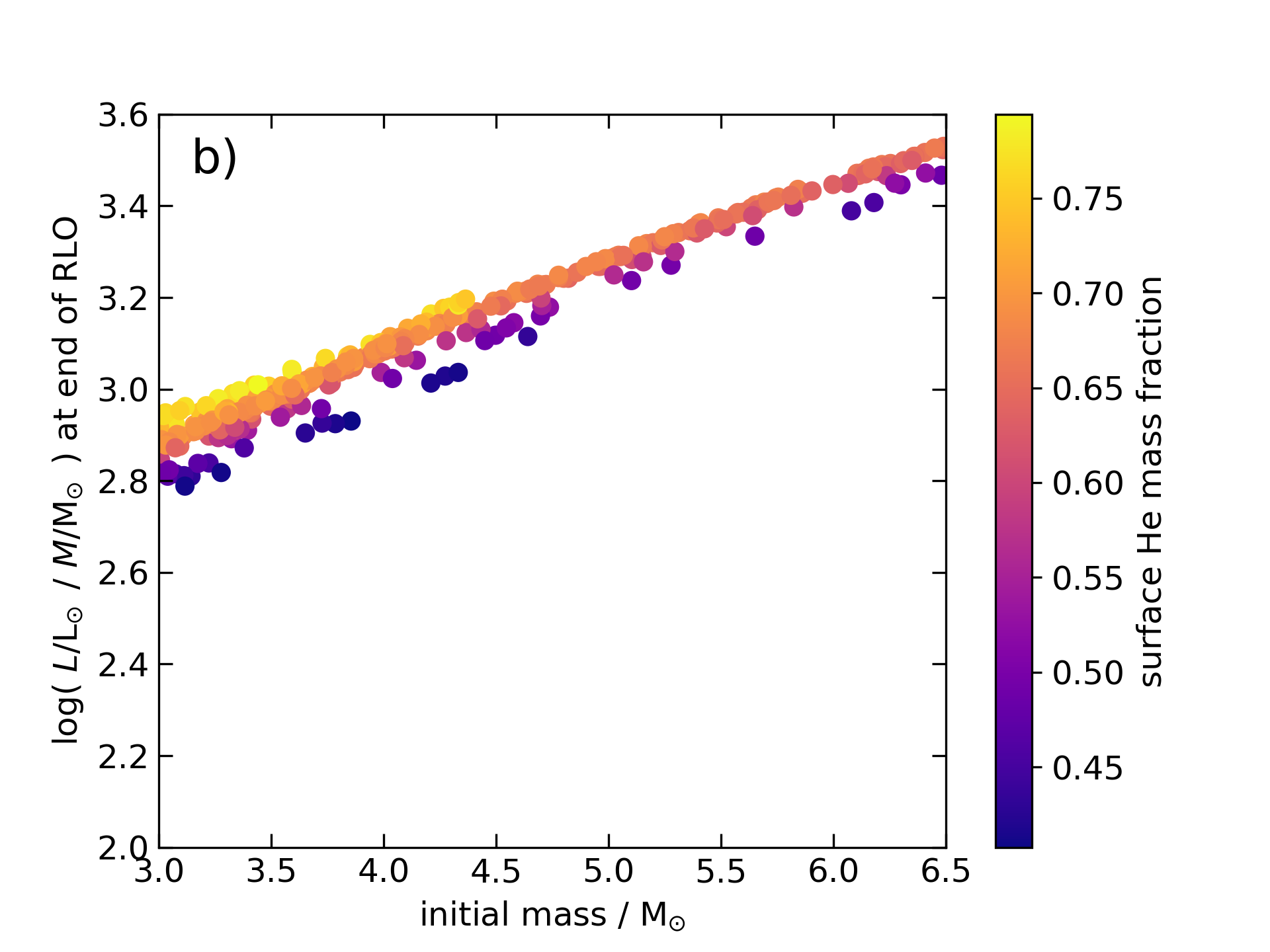}
    \includegraphics[width=0.5\hsize]{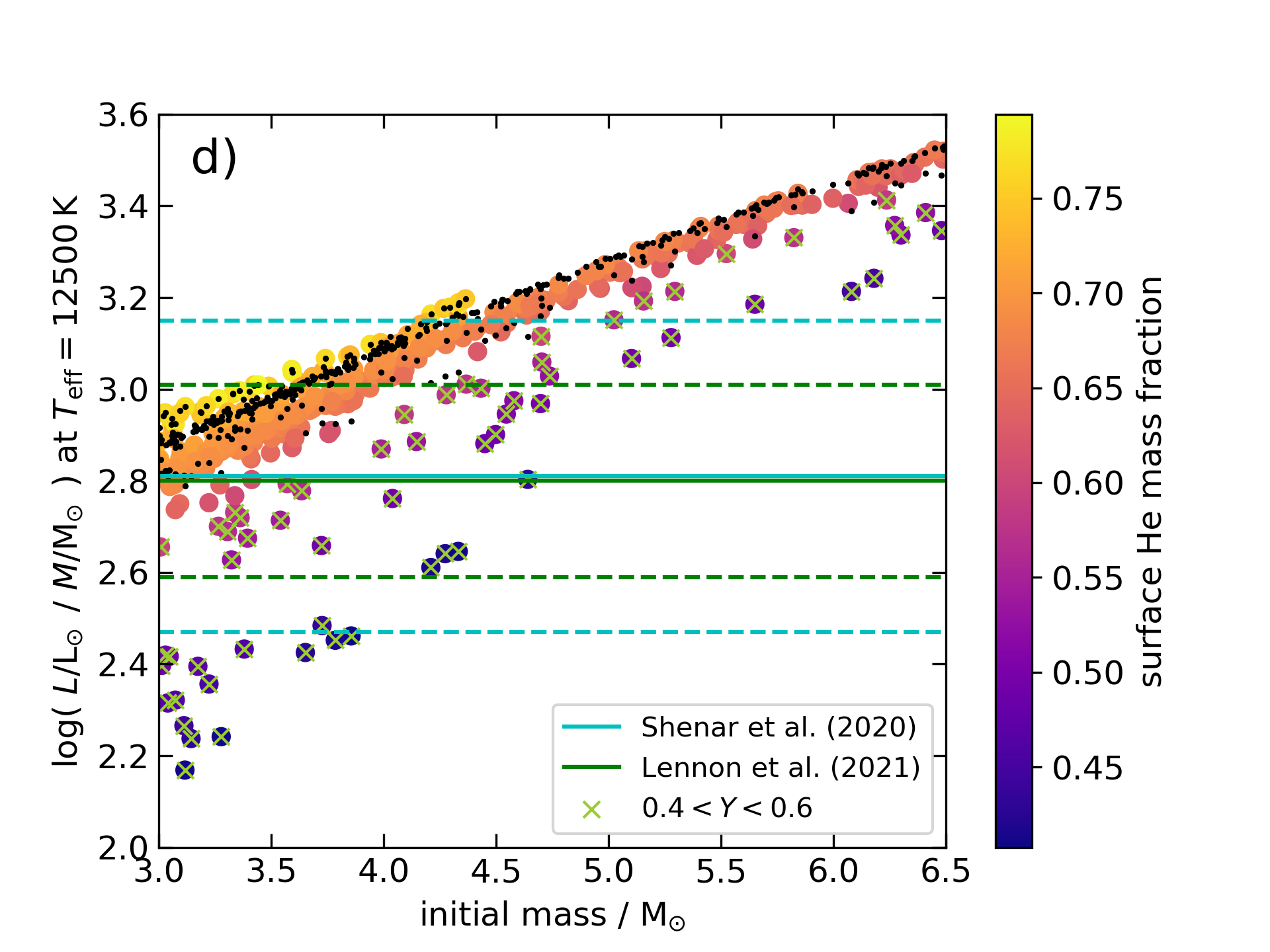}
    \caption{Relation between the initial mass of our models and their luminosity or luminosity-to-mass ratio at selected evolutionary phases.
    {\bf a)} Luminosity and helium core mass of the mass donor immediately after the end of RLO as a function of the initial mass. All shown models underwent a Case~B mass transfer.
    {\bf b)} Luminosity-to-mass ratio and surface helium mass fraction immediately after end of RLO as well as the initial masses of the  models above.
    {\bf c)} Luminosity and surface helium mass fraction when the donor star surface temperature is 12\,500\,K during the contraction after RLO Case~B as a function of initial mass. The black dots represent the state immediately after RLO (panel a). A few models have such a massive envelope that they never reach 12\,500\,K, and are not included.
    {\bf d)} Luminosity-to-mass ratio and surface helium mass fraction when the surface temperature is 12\,500\,K depending in the initial mass. The black dots represent the state at RLO end (panel b). The blue and green lines (almost superimposed) indicate the  $L/M$   from \citet{shenar} and \citet{lennon} with errors shown as dashed lines. Models with a surface helium abundance within the adopted range ($[0.4,0.6]$) are indicated by green crosses.}
    \label{fig:4panel}
\end{figure*}

The typical evolution of such a system has been described in previous works, for example by \citet{1991A&A...241..419P}. We use our fiducial model, which we discuss in detail in Sect.~\ref{sec:res-rot2}, as an illustration (Fig.~\ref{fig:hrd}). Both stellar models start as main-sequence models, until the primary ends core hydrogen burning, ignites hydrogen shell-burning, and starts to expand rapidly. The evolution through the Hertzsprung-gap is halted by the finite size of its Roche lobe \citep[Case~B RLO,][]{1967ZA.....65..251K}. The model starts to lose its hydrogen-rich envelope, and the helium-enriched layers are exposed at its surface. The RLO ends before the helium core is completely revealed. The mass transfer phase ends with the ignition of the helium core-burning. After the end of RLO we obtain a stellar model, whose surface is helium- and nitrogen-enriched ($Y=0.59$ and $\mathrm{[N/C]}=2.47$), but still contains hydrogen. The envelope contracts, which results in an increase in   the model's surface temperature and the surface gravity. The shell source eventually turns off as the envelope is not heavy enough to supply the necessary pressure. This reduces the model's luminosity by almost one order of magnitude. The model settles down slightly to the right of the ZAMS of pure helium stars, where it can be identified as a sdOB~star, and continues burning helium in its core. After central helium depletion it expands again, becoming a helium giant. A second RLO can occur, which increases the surface helium abundance further. Otherwise after RLO the surface helium abundance does not change, as no process is present to dredge up material from the interior onto the surface. The model ends its life as a WD.

Our simulations also contain systems that undergo Case~A RLO \citep[RLO while the donor is burning hydrogen in its core,][]{1967ZA.....65..251K}. These systems are not considered in our analysis as their post-RLO orbits are too narrow ($\lesssim 10\,\mathrm{d}$) to be of relevance for LB-1. We are aware that other studies consider Case~A and discuss this aspect in Sect.~\ref{sec:dis-evo}.

We find that the state of the model at the end of a Case~B RLO is very well defined. The initial mass determines the mass of the helium core and this fixes the luminosity of the hydrogen shell source, which is the dominant source of luminosity at the  end of RLO and during the early phase of contraction. This causes the luminosity-to-mass ratio $L/M$ to remain nearly constant for a wide range of surface temperatures, making it an ideal diagnostic tool as it can be determined from spectroscopic observations via $T_\mathrm{eff}$ and $\log g$. The tight correlation between initial mass, helium core mass, and luminosity of the mass donor immediately after RLO is shown in Fig.~\ref{fig:4panel}a, where we plot these quantities of all our models that survive Case~B mass transfer. We only show models with initial masses up to $6.5\ms$ as higher masses  turn out to be too bright to describe LB-1.

The mass of the stripped star model after RLO is not as strictly correlated to the initial mass as its helium core mass. This can be seen in Figure~\ref{fig:4panel}b, where we show $L/M$ instead of the luminosity $L$ of the stripped model. The higher the mass after RLO, the lower the $L/M$, and the higher the (envelope) mass, the lower  the surface helium abundance. The reason is that when the model loses more mass   during RLO, the deeper layers and thus more helium-rich material is exposed.

As in the Bstr+Be-scenario of LB-1 the mass donor has a temperature of about 12\,500\,K, Fig.~\ref{fig:4panel}c shows the relation between the initial mass and the luminosity of the donor for the time when the models reached that temperature while contracting towards the sdOB~phase. It demonstrates that the luminosity of the stripped star models decreased during this early phase of contraction and that this decrease depends on the surface helium abundance. We point out again that this abundance is a tracer of the mass of the envelope. As in horizontal branch stars, a more massive envelope causes the model to have a lower effective temperature than a model with a less massive envelope for the same conditions in the core. As we select the stellar models by surface temperature, we catch the stars with a more massive envelope at a later phase of contraction where the hydrogen shell is dimmer.

Therefore, if one evaluates the  $L/M$ of a stripped stellar model at a specified temperature (here 12\,500\,K), there are two effects adding scatter to its relation with the initial donor mass (Fig.~\ref{fig:4panel}d). Most of the models still lie on one sequence, but a notable fraction deviates towards a lower $L/M$. We also indicate the $L/M$ value of LB-1 derived by \citet{shenar} and \citet{lennon}. It intersects with the dominant sequence of the simulations just below an initial mass of $3\ms$, our lower mass limit. However, a notable fraction of our models lies within the error range allowing for initial primary masses of more than $5\ms$.

In Fig.~\ref{fig:4panel}d we flag all models that show a helium surface abundance within our adopted range. It yields a sequence of models parallel to the main feature that intersects with the most probable observed $L/M$ value at an initial mass of around $4\ms$. These stripped models also have masses of about $1\ms$, in agreement with the mass estimates of \citet{shenar} and \citet{lennon}. We therefore consider the most likely initial primary mass to be about $4\ms$. The initial secondary mass is  therefore  below this value. This is in agreement with the mass estimate of the Be~star in the analysis of \citet{lennon}, who report $3.4^{+3.5}_{-1.8}\ms$, and \citet{shenar}, who find a spectroscopic mass of $5\ms$, where both studies imply slightly different accretion efficiencies on the secondary. In any case, we found no strong dependencies of the accretor masses on the properties of the stripped models. We  return to the accretion efficiency in Sect.~\ref{sec:dis-evo}, where we also discuss the  implications for the orbital evolution of the system.

\subsection{Post-core-He-burning expansion phase and pre-WD phase}\label{sec:res-other}
So far we have focused on the contraction phase of the donor star immediately after the end of the RLO. However there are two other evolutionary stages during which the donor star may be observationally picked up in the B star regime. These are the expansion phase following core helium depletion, and the transition from the helium giant branch towards the WD phase, which may or may not be separated by a second RLO phase during helium shell-burning \citep[Case BB,][]{1976A&A....47..231S, 1976Ap&SS..43...35D}. In these two evolutionary stages, the stripped star's luminosity in our fiducial model (Fig.~\ref{fig:hrd}) exceeds the value it has during the contraction phase after the first RLO stage by about a factor of~2 and~10, respectively. On the other hand, the lifetime in the B star regime is shorter in these two stages, by factors of~3 and 160 compared to the post-RLO contraction phase. In a population study, certainly the first two of the considered stages should be taken into account, while the pre-WD evolutionary stage is less likely to be observed as it is very short. In this paper we focus on the first stage, since it has the largest observing probability \citep[see also][]{elbadry2} and,  as we discuss  below, for a given effective temperature in the B-type regime the rotational velocities in the first two crossings of
the B-type regime are quite similar (see Sect.~\ref{sec:res-rot2}). 

During the transition from the helium giant branch or a second (Case~BB) mass transfer stage towards the WD phase, the stellar models are typically more than one order of magnitude more luminous than in the contraction phase after the first RLO. This means that if LB-1 was on such a path, its donor ZAMS masses would need to be much lower than $4\ms$, which in in disagreement with the empirical $L/M$ ratio. We therefore consider it unlikely that LB-1 is in that stage.

\subsection{Spin evolution}\label{sec:res-rot}

\subsubsection{Our fiducial model}\label{sec:res-rot2}

%\begin{figure}
%    \includegraphics[width=1.1\hsize]{pic/M4P16-HRD.png}
%    \caption{...}
%    \label{fig:hrd}
%\end{figure}

Since our model grid analysed in Section~\ref{sec:res-pro} was not dense enough for our purpose, we computed an additional model suitable for the analysis of LB-1 according to the findings above. Its initial parameters are ZAMS masses of $4\ms$ and $3.5\ms$, an initial orbital period of $16\days$, and an initial equatorial rotational velocity of 0.6 times the critical velocity. The evolution of the mass donor in the Hertzprung--Russell diagram is shown in Fig.~\ref{fig:hrd}. The model's primary star matches the observed values of $T_\mathrm{eff}$, $L/M$, surface helium, and [N/C] as well as the mass ratio  determined by \citet{shenar}. The final orbital period of $223\days$ is longer than the observed one. The final orbital period is not relevant for the stripped star's properties, and is  discussed  in Sect.~\ref{sec:dis-evo}.

The evolution from Roche-lobe decoupling to the subdwarf stage of this model takes several million years, with a strongly decelerating rate of change of temperature and luminosity. This means that the more a stripped star has contracted, the more likely it is to be observed. The subdwarf phase lasts about 20 million years. Therefore, it is not surprising that we know of almost two dozen  systems with subdwarfs \citep{2018ApJ...853..156W, 2018ApJ...865...76C, 2021AJ....161..248W}, but only three (LB-1, HR~6819, and possibly NGC~1850 BH1) with stripped stars near the main sequence and none to the right of the main sequence.

\begin{figure}
    \includegraphics[width=\hsize]{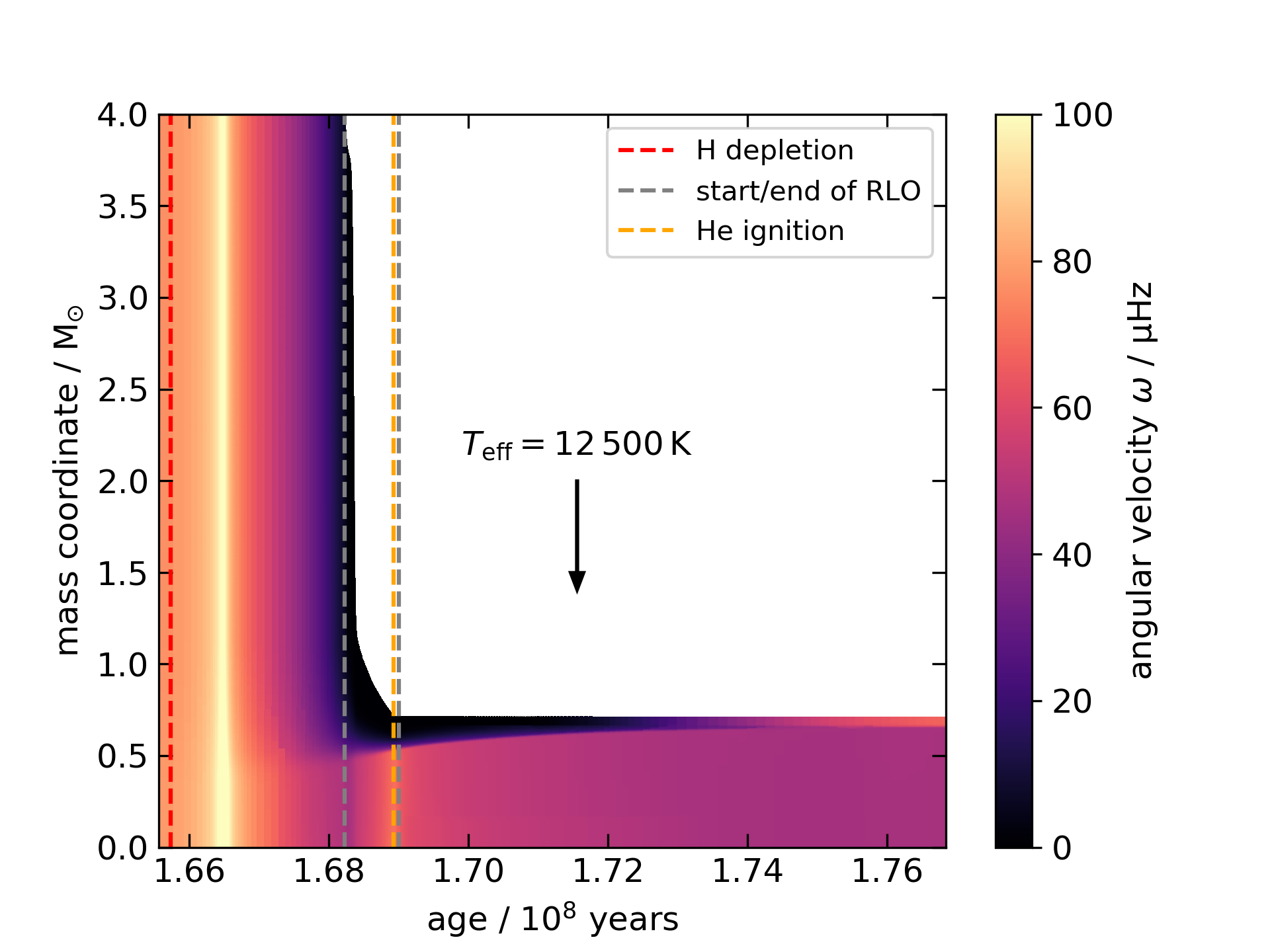}
    \includegraphics[width=\hsize]{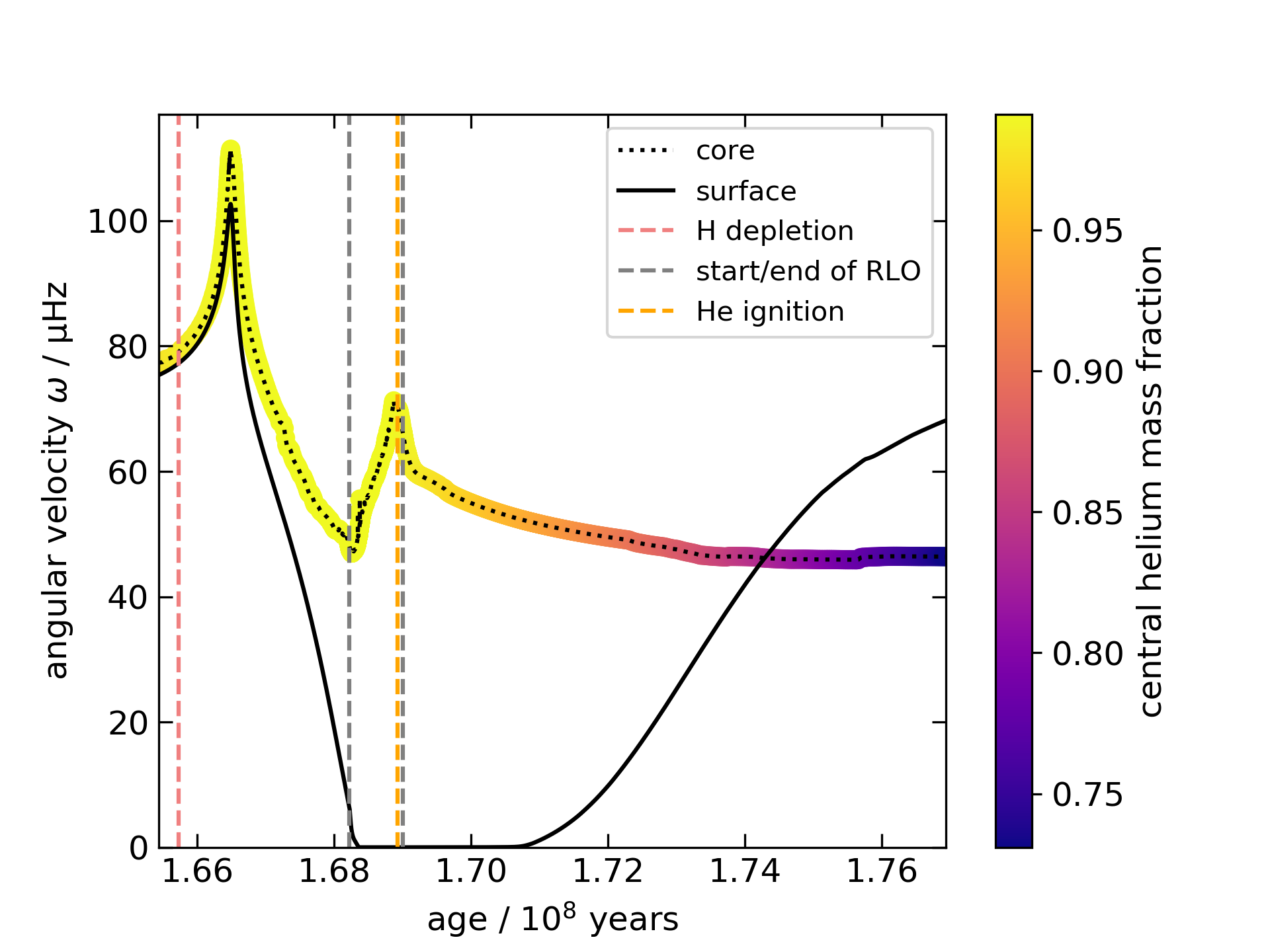}
    \caption{Evolution of the rotation frequency of the mass donor of our fiducial model.
    Top: Kippenhahn-type diagram the model, showing the internal evolution of its rotational frequency (see colour bar at right) after TAMS. The arrow marks the time when $T_\mathrm{eff} = 12\,500\,\mathrm{K}$.
    Bottom: Angular velocity of the model near its centre, colour-coded by the central helium mass fraction, and at the surface. The vertical dashed lines in both panels indicate certain evolutionary steps.}
    \label{fig:kippen}
\end{figure}

The evolution of the internal rotation of our fiducial mass donor is depicted in Fig.~\ref{fig:kippen} (top). Until the end of core hydrogen burning, the model rotates close to a rigid body. At terminal age main sequence (TAMS), at an age of about $1.665\cdot10^8\,$years, core and envelope have an angular velocity of roughly $100\,\text{\textmu Hz}$. Thereafter the envelope expands, the core contracts, and eventually their rotation rates grow apart. This can be seen in Fig.~\ref{fig:kippen} (bottom) where we show the rotation frequency of the innermost mass shell and the surface of the model. The difference between the core and surface rotation rates increases with time indicating that the rotational coupling weakens. The surface slows down to a velocity of about $10^{-2}\kms$ due to tides and its growing moment of inertia and adjusts to the binary orbital frequency of about $0.2\,\mathrm{nHz}$. On the other hand, the core halves its rotation rate due to the combined effect of the Spruit--Tayler dynamo, which decelerates the core, and contraction of the helium core, which accelerates it.

The rotation of core and envelope are decoupled at an age of $1.68\cdot 10^8$ years; at this age  the core is not slowing down, but rather is increasing due to contraction,  its rotation rate uninhibited by the slowly rotating envelope. This time coincides with the age when the model fills its Roche lobe. The core rotation rate reaches a maximum at an age of about $1.69\cdot 10^8$ years, which is when the RLO ends and helium core-burning is ignited. The end of the RLO and helium ignition take place nearly at the same time since helium ignition terminates the core contraction, and thus the envelope expands due to the mirror principle \citep{1967ZA.....65..251K}. Only a small fraction of the envelope remains at this time, which is still rotating at the same frequency as the orbit. What follows is the contraction of the envelope, speeding up its rotation. The rotation of the core does not change much  more thereafter. Its mass grows as long as the hydrogen shell-burning is active, and thus low angular momentum material is incorporated into the core.

\begin{figure}
    \includegraphics[width=\hsize]{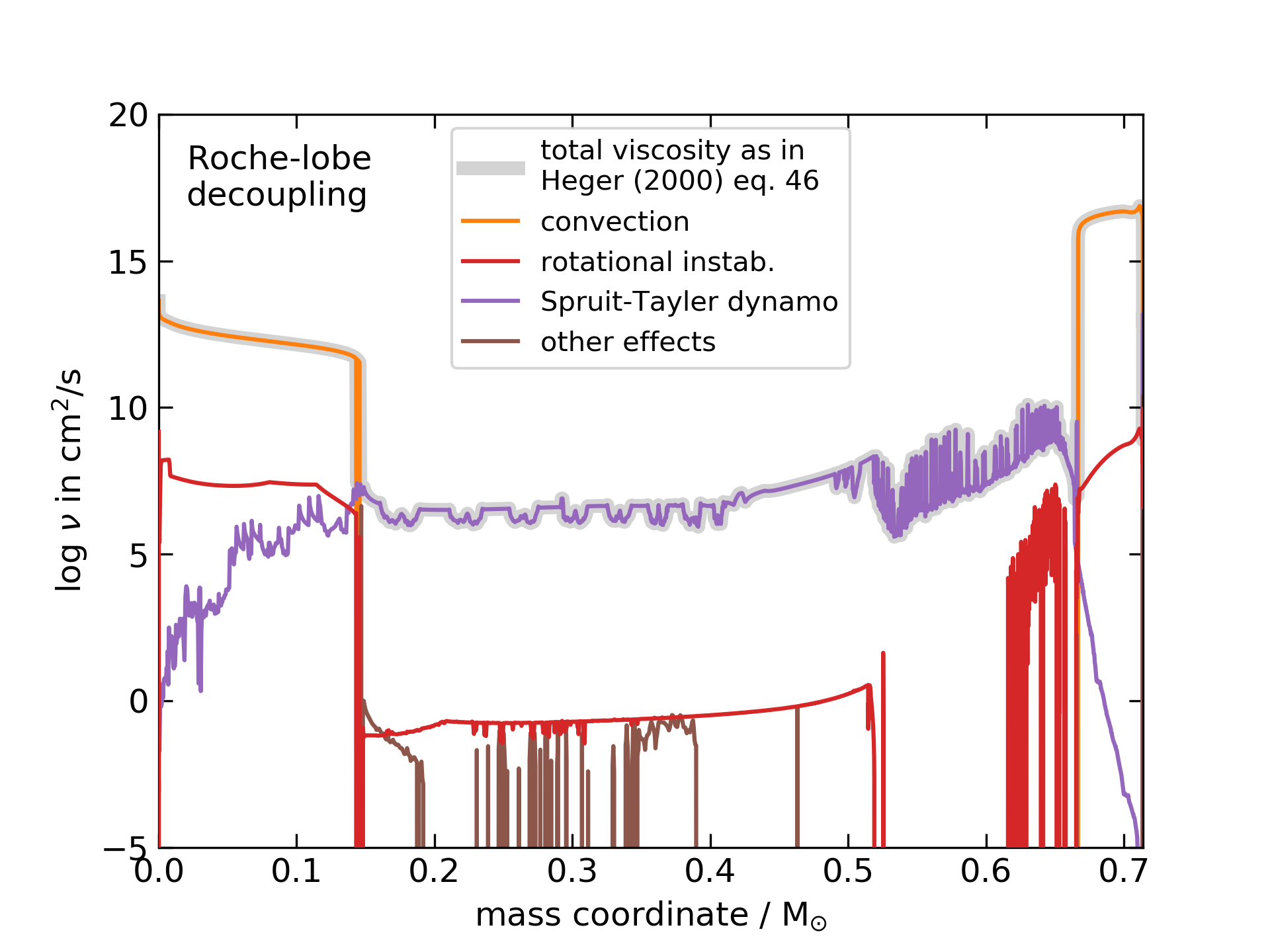}
    \includegraphics[width=\hsize]{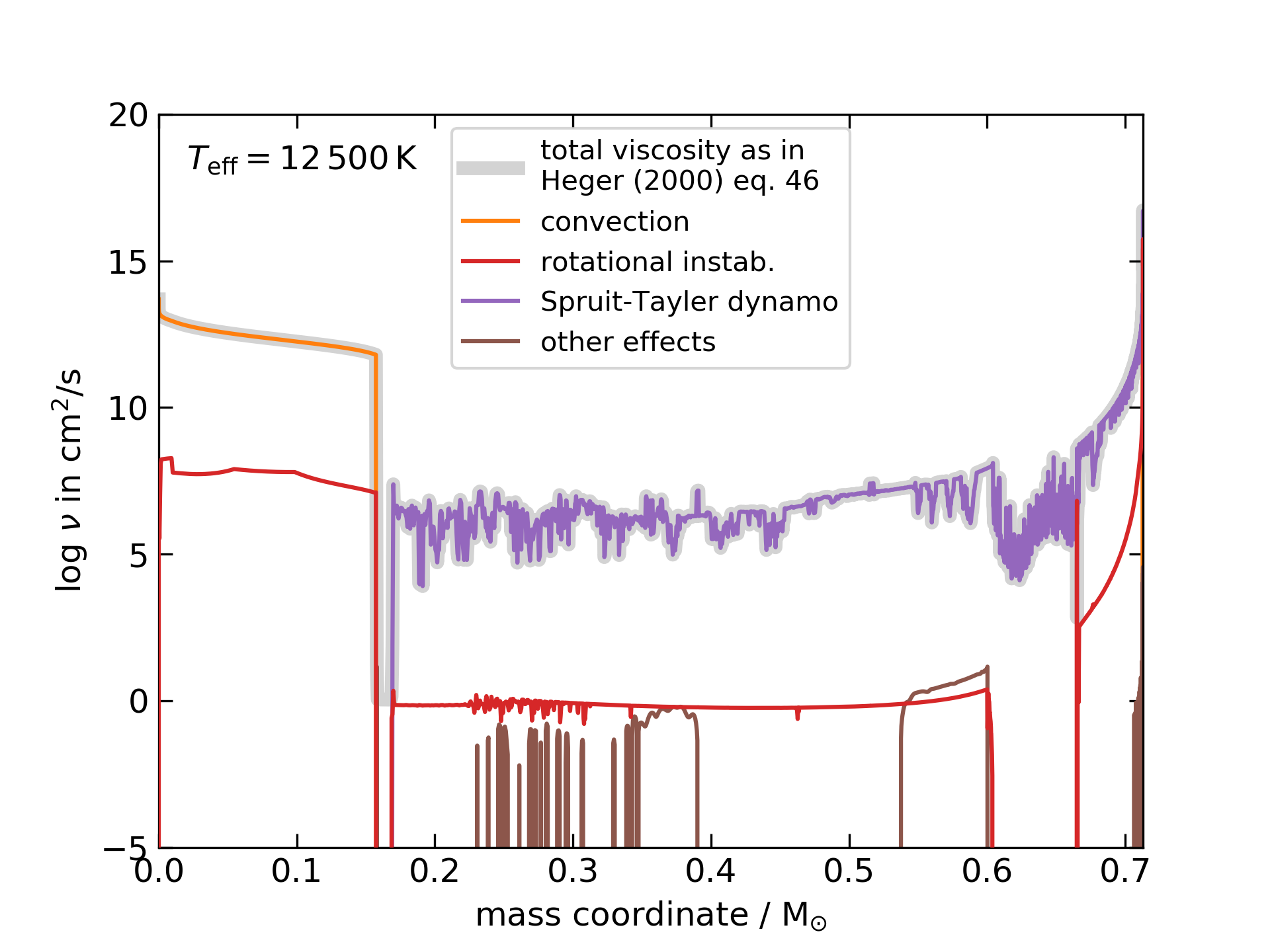}
    \caption{Contributions to the viscosity~$\nu$ for the angular momentum transport as a function of mass coordinate in our fiducial model, at the time of Roche-lobe decoupling (top) and when the effective temperature reached 12\,500\,K (bottom).}
    \label{fig:visc}
\end{figure}

We have thus shown that during core hydrogen burning the star rotates close to a rigid body and that after central hydrogen exhaustion, the rotation rates of core and envelope decouple. This can be explained by the magnetic torque of the Spruit--Tayler dynamo. During hydrogen core-burning, the gradients of entropy and mean molecular weight are small enough to result in a high magnetic viscosity, which maintains core and envelope at the same rotation rate. This is not true any more after central hydrogen exhaustion, when the hydrogen shell-burning enlarges these gradients. This can be seen in Fig.~\ref{fig:visc}, where we show the internal profiles of the effective viscosity and its contributions. A clear drop is visible at mass coordinates of $\sim 0.52\ms$ (top) and $\sim 0.61\ms$ (bottom). This drop in the magnetic viscosity makes the angular momentum transport between core and envelope so inefficient that they develop different rotation rates. Convection only plays a role in the helium-burning core and in the outer envelope during early contraction (orange lines in Fig.~\ref{fig:visc}). The helium burning core and the radiative zone above, which together form the helium core, are strongly coupled as predicted by \citet{2014ApJ...793..123M}.

At an age of $\sim1.74\cdot10^8$ years the envelope starts to move more quickly than the core, leading to a positive angular velocity gradient in the early sdOB~phase. This occurs due to the rapid contraction of the envelope after the end of RLO, from a convective expanded state to a radiative and more compact structure \citep{1998A&A...334..210H}. This feature demonstrates that core and envelope evolve independently, meaning that the evolution of the rotation rate of the envelope is not driven by the core, but rather by its own contraction. The $\del \omega / \del m$ term in Eq.~\ref{eq:dodt} would not lead to a deviation from uniform rotation, which was nearly archived at an age of $\sim1.74\cdot10^8$ years. During the helium core- and shell-burning phases the core and envelope continue to rotate at different angular velocities. The rotation rates do not adjust since no angular momentum exchange between core and envelope takes place (i.e. they remain rotationally decoupled).

\begin{figure}
    \includegraphics[width=1.1\hsize]{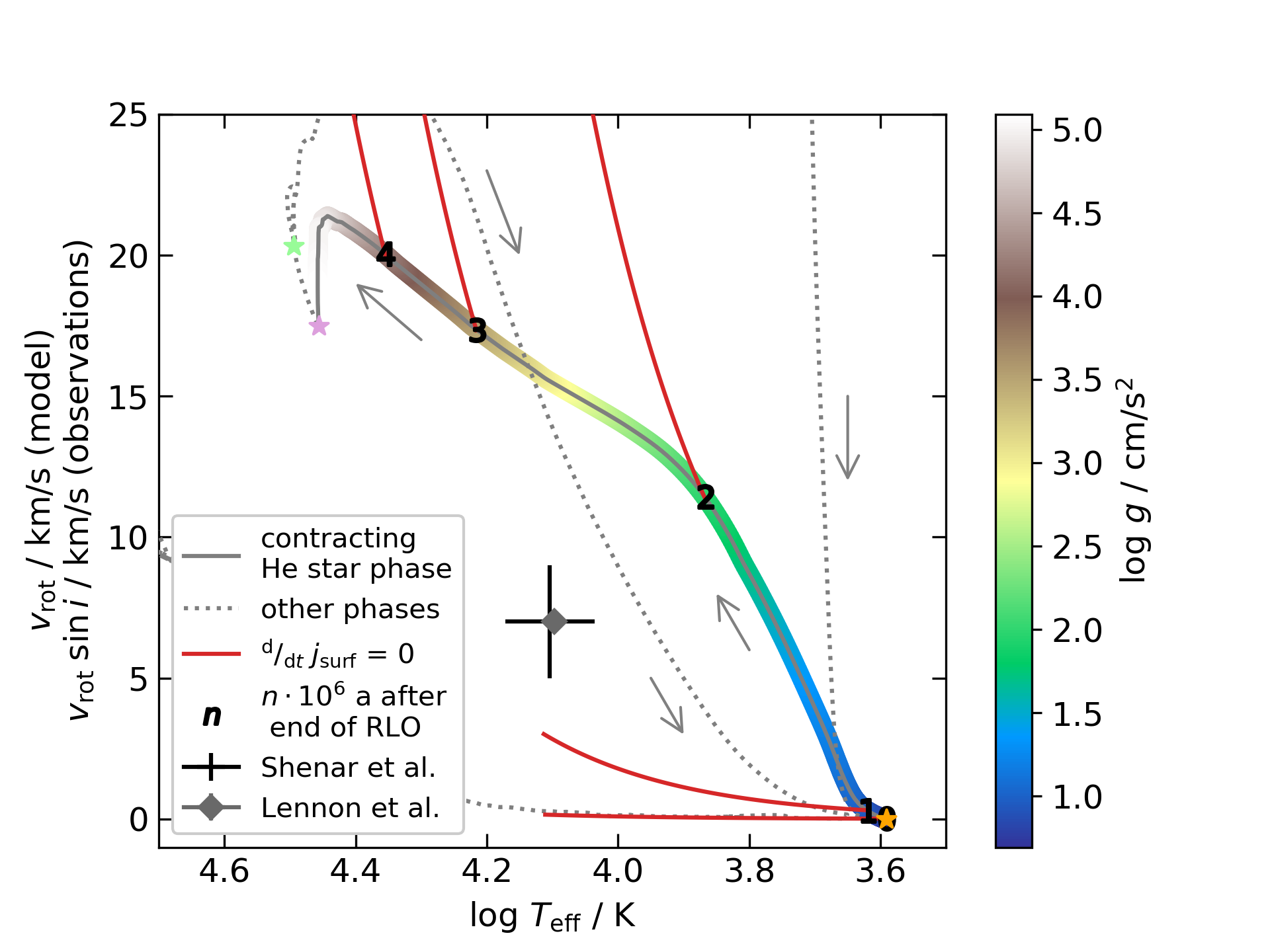}
    \caption{Surface rotational velocity and surface gravity (colour-coded) as a function of the effective temperature of the stripped component of our fiducial binary model. The coloured part of the curve indicates the contraction phase from RLO decoupling towards the sdOB~phase. The dashed black lines show the evolution before and after that phase, with arrows indicating the direction of evolution. The time elapsed after the RLO is indicated in Ma by the numbers on the curve. The two observational data points (black and grey) are  from \citet{shenar} and \citet[][the errors are smaller than symbol size]{lennon}. The measured rotational velocity includes a factor of $\sin i$. The red lines indicate the rotational evolution if the specific angular momentum of the stellar surface was conserved. The star symbols indicate the  evolutionary stages, as in Fig.~\ref{fig:hrd}.}
    \label{fig:Tv}
\end{figure}

Figure~\ref{fig:Tv} shows the evolution of the surface rotational velocity of our fiducial model from the end of RLO until the end of the contraction towards the sdOB~state. In this phase the model's radius decreases, while the  mass and angular momentum remain constant. As the star shrinks, the surface temperature, gravity, and rotational velocity increase. While the evolution of temperature and gravity can be simply explained by the change in radius assuming constant luminosity, the rotational evolution cannot, as indicated by the red lines arising from the main curve every one million years in Fig.~\ref{fig:Tv}. They reflect the evolution of the rotational velocity under the assumption that the specific angular momentum at the stellar surface is conserved, that is 
\begin{equation}\label{eq:vT}
    \vrot = R_0 {\vrot}_{,0} \sqrt{\frac{4\pi\sigma T_\mathrm{eff}^4}{L}\,},
\end{equation}
where $R_0$ and ${\vrot}_{,0}$ are the radius and the rotational velocity at that point where the red lines start (i.e. every million years).

The first two red lines follow lower rotational velocities than the model curve, which implies that the surface layers in the MESA model gain angular momentum, which  happens because the tidal forces do not keep the surface in co-rotation any more. Because they behave as $(R/a)^6$ \citep{1977A&A....57..383Z}, a small decrease in the radius can render the tides inefficient. Hence, the convective angular momentum transport forces the envelope back to rigid rotation (see Fig.~\ref{fig:kippen}). Thereafter the contraction of the envelope leads to a reduction of the surface angular momentum. This can be seen by the red lines lying \emph{above} the evolutionary curve. The physical reason is that the envelope rotates almost as a rigid body (see Fig.~\ref{fig:kippen}) and that its angular momentum is conserved. As the radius of the core envelope boundary, and therefore the moment of inertia of the base of the envelope, barely change after helium ignition \citep{2013sse..book.....K}, the contraction induced rigid acceleration of the envelope increases the specific angular momentum of the base of the envelope. This angular momentum can only come from the top of the envelope, and thus the surface's specific angular momentum decreases \citep[see also][]{1998A&A...334..210H}.

Figure~\ref{fig:Tv} also shows the measurement of \citet{shenar} and \citet{lennon} for the stripped star in LB-1. While both numbers lie below the model curve, they include a factor of $\sin i$. With an inclination of approximately $30^\circ$ ($\sin i \approx 0.5$), which is close to the inclination value inferred by \citet{shenar}, our model consistently fits the observed properties of LB-1 in the Be+Bstr-scenario. \citet{lennon} argue for $\sin i \approx 0.8$, but (as they note) their spectroscopic mass estimate assumes spherical symmetry.

\subsubsection{Alternative models}\label{sec:res-alt}
To study the influence of the initial stellar spin, we computed another model for which we kept all parameters as in our fiducial model, except for the initial rotation velocity, which we fix here at 0.2 times the critical rotation at ZAMS. We then computed two additional models identical to the fiducial one and its slower spinning counterpart, but neglecting magnetic angular momentum transport. The figures corresponding to the three alternative models (\ref{fig:20vvc}, \ref{fig:noB}, and \ref{fig:20noB}) can be found in the Appendix.

Setting the initial rotational velocity to $20\%$ of the critical velocity reduces the angular velocities in the whole the star during central hydrogen burning (Fig.~\ref{fig:20vvc}). After RLO, the core spins notably more quickly than before, with a rotation rate nearly identical to that of the fiducial model at that time (Fig.~\ref{fig:kippen}, bottom). The reason is that core and envelope decouple earlier than in the fiducial model, namely between TAMS and Roche-lobe filling. Hence, the core transfers less angular momentum to the envelope. The envelope rotation evolves similarly to the original model (Fig.~\ref{fig:Tv} and \ref{fig:20vvc} lower left panel). The rotation of the envelope is controlled by the same tidal forces, which set the surface velocity to the same synchronised value in both cases. Therefore, the surface rotation during the contraction phase afterwards also occurs as in the fiducial model.

Turning off the magnetic angular momentum transport has strong effects on the rotational evolution, independently of the initial spin. This is visible in Figs.~\ref{fig:noB} and \ref{fig:20noB}, which show that core and envelope already decouple during the central hydrogen burning phase. This leads to a core spinning much faster   than the envelope as the former contracts and the latter expands on a timescale shorter than the timescale of any present angular momentum transport process. At the end of RLO the envelope is rotating extremely slowly, as in the original model, but is soon spun up by the core, even though magnetic coupling is absent. The angular velocity gradient is large enough to transport angular momentum from the core to the envelope even if only the hydrodynamic rotational instabilities contribute to the viscosity. However, unlike in the magnetic case, a strong angular velocity gradient between core and envelope remains. Comparing Fig.~\ref{fig:kippen} (bottom) and Fig.~\ref{fig:noB} (bottom right) reveals the impact of the magnetic coupling;  before the onset of helium burning the core's angular velocity grows at a much larger rate than in the fiducial model. Figure~\ref{fig:noB} (lower left panel) shows that the non-magnetic simulation with an initial rotation of 0.6 times the critical rotation is not in agreement with LB-1.

The model without the dynamo and with an initial rotation rate of 0.2 times the critical rotation (Fig.~\ref{fig:20noB} bottom left), however, is in agreement with LB-1. During the contraction phase the model's rotational velocity is about $10\kms$, which fits with LB-1. When the model has ended its contraction and reached the sdOB~phase, the rotational velocity is about $50\kms$. This value is caused by the fast rotation of the core as the strong angular velocity gradient between core and envelope together with the hydrodynamical angular momentum transport mechanisms increase the rotation of the envelope notably (Fig.~\ref{fig:20noB} bottom right), and in contradiction with the empirical rotational velocities of sdOB~stars like \textphi~Per discussed in Sect.~\ref{sec:dis-obs}. All the observed sdOB~stars show rotational velocities below $50\kms$ (Sect.~\ref{sec:dis-obs}), except one (QY Gem). This means that only stellar models with the Spruit--Tayler dynamo can reproduce the observed rotational velocities after a RLO. The initial rotation rate does not play a role.

\subsubsection{Comparison to single-star models}\label{sec:res-sing}

\begin{figure}
    \includegraphics[width=\hsize]{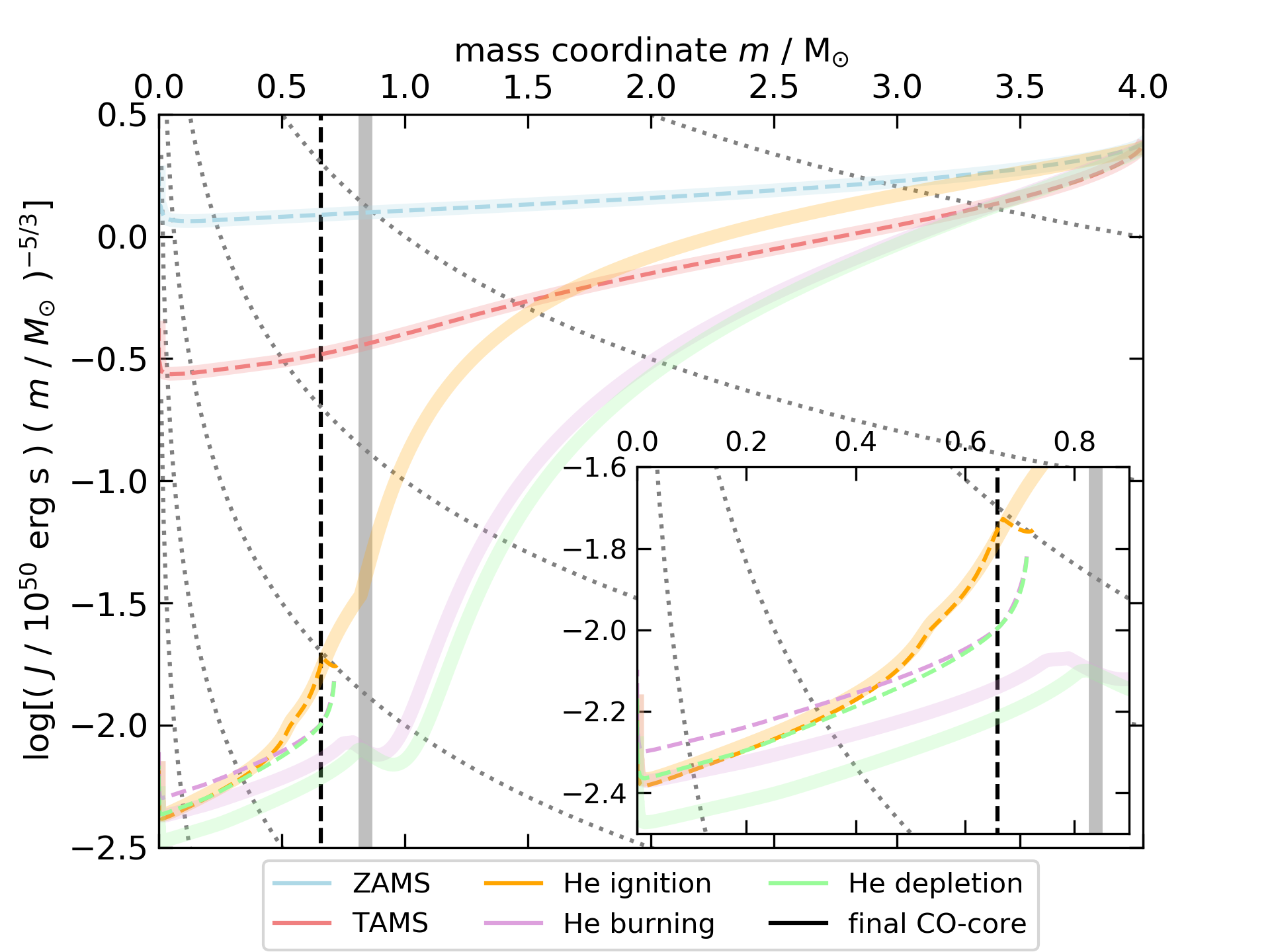}
    \includegraphics[width=\hsize]{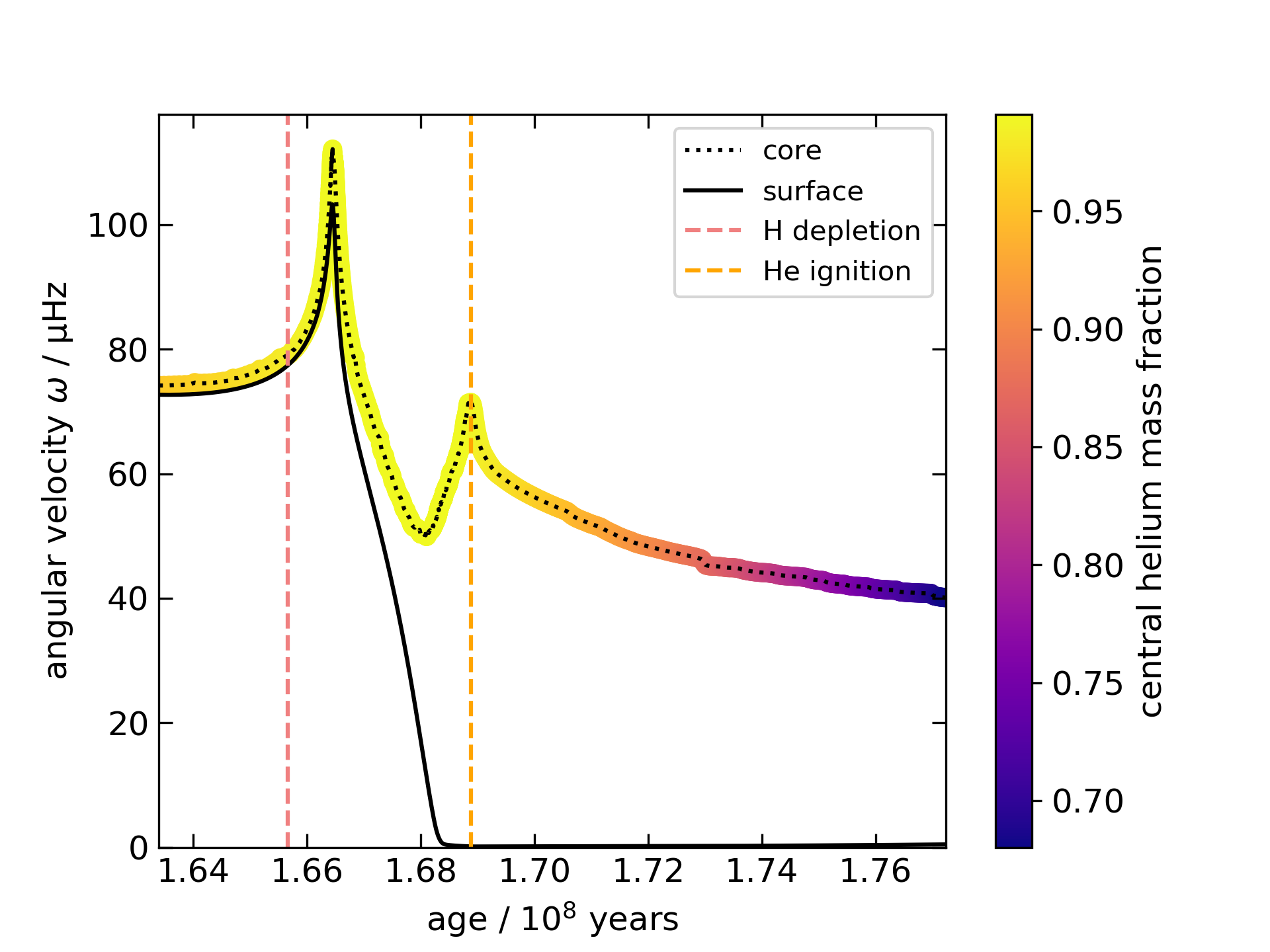}
    \caption{Rotational properties of the single-star model.
    Top: Angular momentum $J(m)$ within the mass coordinate $m$ divided by $m^{5/3}$ as a function of $m$ at different evolutionary stages for the primary star of the binary model examined in Sect.~\ref{sec:res-rot2} (dashed) and its primary star as a single star (thick lines). The vertical black lines indicate the last calculated mass of the CO-core of the models. The thin dotted lines represent constant values of the angular momentum $J$: (from top to bottom) $\log J/10^{50}\,\mathrm{erg\,s} = \{1,0,-1,-2,-3,-4\}$.
    Bottom: Evolution of the angular velocity of the single-star model near the centre (coloured by the central helium mass fraction) and at the surface. The vertical lines indicate hydrogen depletion and helium ignition.}
    \label{fig:sing}
\end{figure}

Because of its relatively low mass, the stripped star of LB-1 will end up as a WD. The same is true if the stripped star's progenitor is a single star. Here, we investigate whether the binary interaction has an influence on the WD spin. Figure~\ref{fig:sing} (top) displays $J(m)/m^{5/3}$, which is the integrated angular momentum $J(m) = \int_0^m j(m) \,\mathrm{d} m$ at mass coordinate $m$, divided by $m^{5/3}$, such that homogeneous and rigidly rotating regions would result in a horizontal line in the plot \citep{2008A&A...481L..87S}. We plot this quantity at ZAMS, TAMS, helium ignition, in the middle of helium burning when the luminosity has its lowest value, and at helium depletion, for both the $4\ms$ star in the binary discussed in Sect.~\ref{sec:res-rot2} and as a single-star model. The lines from different evolutionary stages trace the flow of angular momentum through the mass shells since $J(m)$ and therefore $J(m)/m^{5/3}$ remain constant in a given mass shell if no angular momentum is transported through this shell.

At ZAMS, TAMS, and helium ignition there is no significant difference in the angular momentum distribution between the single-star and binary model (except that the missing envelope in the star of the binary model results in a lower mass and that the Roche lobe filling star holds barely any angular momentum in its outermost layers). For the first two evolutionary stages this is not surprising as the binary model behaves like a single-star model if it remains much smaller than its Roche lobe. Flat curves indicate rotation close to a rigid body. The curve is slightly steeper at TAMS than at ZAMS since angular momentum flows from the core into the envelope \citep{2008A&A...481L..87S}. During the transition from TAMS to helium ignition the core rotation decouples from that of the envelope. Figure~\ref{fig:sing} (bottom) shows that this happens at the same time in the single-star model as in our fiducial binary model (Fig.~\ref{fig:kippen}). The evolution of the core's angular velocities in the two models is nearly identical during the shown times. Likewise the angular momentum distribution at helium ignition of the binary model and in the core of the single-star model are almost indistinguishable. This indicates that at this stage it does not matter whether the rotation of the envelope is braked by tidal forces or by the expansion induced by the variation of the moment of inertia. The strong gradient in the angular momentum profile shows that the models are rotating differentially and that in the case of the  single-star model angular momentum was transferred from the core to the envelope.

Slight differences in the angular momentum distribution between the single-star and binary model appear during helium burning. In the model of the stripped star some angular momentum is lost due to tides when the star still fills a notable fraction of its Roche lobe. This can be seen in Fig.~\ref{fig:sing} (top), where the orange dashed line (helium ignition) extends to larger angular momenta than the purple line (middle of helium core-burning) and the green line (central helium depletion). Thereafter the angular momentum of the stripped star model is conserved and only subtle changes of the internal angular momentum distribution occur (the ends of the green and purple dashed lines overlap). In the single-star model, hydrogen shell-burning increases the core's mass and angular momentum, but decreases its specific angular momentum as the added envelope material contains little angular momentum. This is indicated by the green line, which lies  below the purple line, which in turn  lies below the orange line. When material is added to the core, it adapts to the core's rotation, as shown in Fig.~\ref{fig:sing} (top). The bumps in the thick green and plum line around $m=0.8\ms$, which indicate the angular velocity gradient between core and envelope, move to a higher mass coordinate. This happens because below the burning shell no chemical gradient weakens the magnetic torques.

To estimate the spin angular momentum of the emerging WD, we use the spin of the CO-core at helium depletion as no significant changes in the core specific angular momentum are expected in later phases \citep{2008A&A...481L..87S}. The binary model was terminated during the contraction phase following central helium exhaustion, and the evolution of the single-star model ended during the thermally pulsing asymptotic giant-branch phase. We use the CO-core masses of the last calculated model of the simulations to estimate the expected WD masses, which differ due to hydrogen  shell-burning in the single-star case, as $0.66\ms$ for the binary model and $0.84\ms$ for the single star. The integrated angular momenta at this mass coordinate at the time of central helium depletion is equal to $5.03\cdot10^{47}\,\mathrm{erg\,s}$ for the binary model and to $5.82\cdot10^{47}\,\mathrm{erg\,s}$ for the single-star case. The corresponding mean specific core angular momenta are $3.82\cdot10^{14}\,\mathrm{cm^2/s}$ (binary) and $3.47\cdot10^{14}\,\mathrm{cm^2/s}$ (single). Thus, we find no significant imprint of the binary evolution on the resulting WD spin. This is not surprising as we have already shown that cores evolve nearly identically until helium ignition in the two models (Fig.~\ref{fig:sing}). After that the rotation of the core of the binary model is decoupled from the envelope and the changes in the core's mass and angular momentum of the single-star model are small.

\subsection{Rotation, orbital period, and helium abundance after RLO}\label{sec:res-prog}

\begin{figure}
    \includegraphics[width=1.1\hsize]{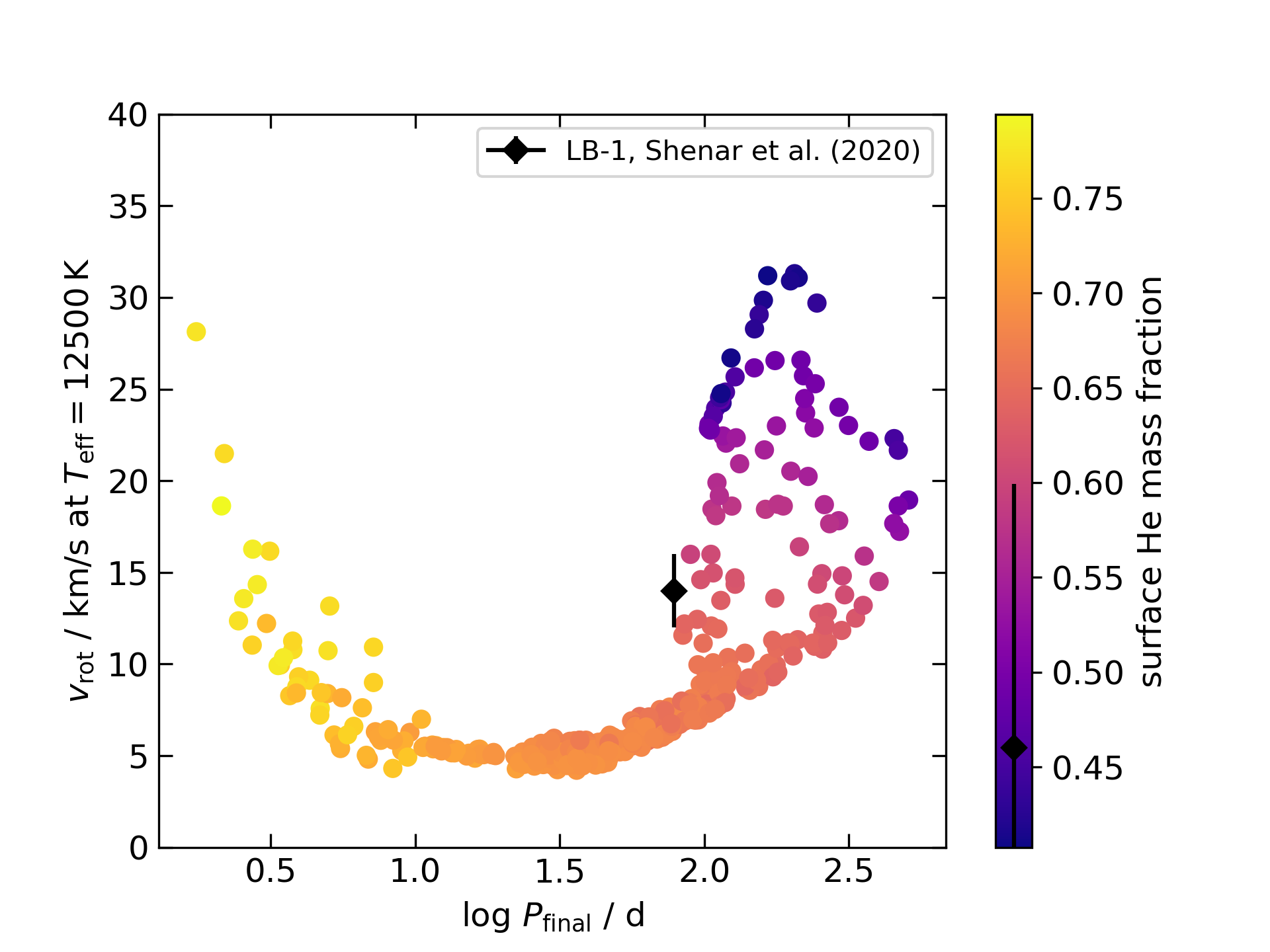}
    \caption{Orbital period vs donor rotational velocity, and surface helium abundance (colour-coded) of our donor star models contracting after RLO when reaching an effective temperature of 12\,500\,K. The corresponding quantities derived for LB-1, assuming an inclination of $ i = 30^\circ$, are shown by the black dot \citep{shenar}.}
    \label{fig:prog}
\end{figure}

Here we consider the predicted properties of stripped star models at  given effective temperatures during their contraction, using the set of stellar models discussed in Sect.~\ref{sec:res-pro}. Figure~\ref{fig:prog} shows the orbital period and donor surface helium abundances after RLO together with the rotational velocity of the donor. While the orbital period and surface abundances remain constant during the contraction phase, the rotational velocity increases (see Sect.~\ref{sec:res-rot2}). Figure~\ref{fig:prog} depicts the equatorial rotational velocity of our models at the time when their effective temperature has reached 12\,500\,K. 

The relation between orbital period and rotation follows a narrow U-shape for two reasons. For short orbital periods tidal forces keep the donor in co-rotation, and longer periods imply smaller rotation velocities. For wide binaries, the donor radius at Roche lobe decoupling is large. Hence the star must contract further to reach the adopted surface temperature, which leads to larger rotational velocities. 

For periods above $\sim$100\,d, the distribution in Fig.~\ref{fig:prog} widens in two ways. At a given period, the models show a wide range of rotation rates, and a wide range of surface helium abundances, such that the faster rotators show a smaller helium abundance. We show  in Sect.~\ref{sec:dis-evo} and Fig.~\ref{fig:PeHe} that this behaviour can be attributed to the onset of convection in the donor during RLO. 

For Fig.~\ref{fig:prog} we adopted a fixed primary effective temperature of 12\,500\,K. As we  show in Sect.~\ref{sec:res-rot2}, the surface angular momentum is not conserved during contraction. Therefore, the product ${\vrot}{T_\mathrm{eff}^{-2}}$ is not constant (cf. Eq.~\ref{eq:vT} for constant luminosity), and the distributions for different values of the donor surface temperature do not follow in a straightforward way from Fig.~\ref{fig:prog}. We provide plots for the different temperatures in Fig.~\ref{fig:prog2}.

As mentioned above, our models predict a longer orbital period than that observed in LB-1, which may be related to the low mass transfer efficiency of our models (see  Sect.~\ref{sec:dis-evo}). Higher mass transfer efficiencies would shift the models to smaller orbital periods \citep{soberman}. The helium abundance is in broad agreement with LB-1, as the observations are  uncertain. However, we expect that the shape of the distribution of points in Fig.~\ref{fig:prog}, and the connection to the helium abundance, remains qualitatively similar for different mass transfer prescriptions because the physical effects leading to it (tides and contraction) are still in place.  Therefore, observing a large population of contracting post-RLO stars and comparing them with Fig.~\ref{fig:prog} and Fig.~\ref{fig:prog2} would yield strong constraints on the mass transfer efficiency and orbital evolution during RLO.

\section{Discussion}\label{sec:discuss}
In this section we compare LB-1 with the similar system HR~6819, and with sdOB~stars like \textphi~Per (Sect.~\ref{sec:dis-obs}) and relate our results to earlier work on angular momentum transport in stars (Sect.~\ref{sec:dis-calc}). Then we discuss the implications of our results for the orbital evolution of binaries in Sect.~\ref{sec:dis-evo}.

\subsection{Similar systems and \textphi~Per stars}\label{sec:dis-obs}
The system HR~6819 (Table~\ref{tab:atmos_hr}) is resembles to LB-1 in several respects. Similarly to LB-1, it was first suggested that HR~6819  contains a BH since \citet{rivinius} identified it as consisting of a close B3\,III+BH binary with a Be-star orbiting it. \citet{bodensteiner} and \citet{elbadry2} pointed out that it may actually be composed of a Be star and a stripped star. Both studies performed an atmospheric analysis of the stripped star and received consistent results. The stripped component of HR~6819 is slightly hotter and has a stronger surface gravity than that of LB-1. \citet{bodensteiner} derive $\vrot\sin i < 25 \kms$, and \citet{elbadry2} $\vrot\sin i <20\kms$. Both upper limits, imply a high probability for very slow rotation, atypical of B stars, but expected for stripped star (see  Fig.~\ref{fig:prog2}). The Roche-lobe filling component of HD~15124 rotates critically as expected \citep{2022arXiv220105614E}. For NGC~1850 BH1 \citep{2022MNRAS.511L..24E} no rotation rates have been measured so far.

\begin{table}
    \centering
    \caption{Orbital and atmospheric properties of the stripped star in HR~6819 according to \citet{bodensteiner} and \citet{elbadry2}.}
    \label{tab:atmos_hr}
    \begin{tabular}{lcc} \hline\hline
         & \citeauthor{bodensteiner} & \citeauthor{elbadry2} \\ \hline
        $P_\mathrm{orb}\,\mathrm{/d}$ & $40.335\pm0.007$ & $40.3\pm0.3$ \\
        $q$ & $15\pm3$ & $14\pm6$ \\ \hline
        $T_\mathrm{eff}\,\mathrm{/K}$ & $13\,000\pm1\,000$ & $16\,000\pm1000$ \\
        $\log g \,\mathrm{/\,cm/s^2}$ & $2.8\pm0.2$ & $2.75\pm0.35$ \\
        $\vrot \sin i \,\mathrm{/\,km/s}$ & $<25$ & $<20$ \\
        $\log (L/M)$ & $3.41\pm0.23$ & $3.46\pm0.37$ \\
        $Y_\mathrm{surface}$ & $\sim$ solar & $0.54\pm0.11$ \\
        $[\mathrm{N}/\mathrm{C}]$ & $>0$ & $1.6\pm0.4$ \\ \hline
    \end{tabular}
\end{table}

\begin{table*}
    \centering
    \caption{Properties of the sdOB~stars in known and candidate Be+sdOB systems as well as HR~6819.}
    \begin{tabular}{ccccccc}\hline\hline
         HD Number & Name & $T_\mathrm{eff}\,\mathrm{/kK}$ & $\log g\,\mathrm{/cm\,s^{-2}}$ & $\vrot\sin i\,\mathrm{/\,km/s}$ & Orbital period /d & Reference \\ \hline
         10516 & \object{\textphi~Per} & $53\pm3$ & $4.2\pm0.1$ & $<10$ & 127 & \citet{1998ApJ...493..440G} \\%P=127
         29441 & \object{V1150~Tau} &    $40.0\pm2.5$ &   --\tablefootmark{(a)} &    $<15$ &   -- & \citet{2021AJ....161..248W}\\
         41335 & \object{HR~2142} & $>43\pm5$ & $>4.75$ & $<30$ & 80.9 & \citet{2016ApJ...828...47P} \\%P=80.9d
         43544 & \object{HR~2249} &    $38.2\pm2.5$ &   --\tablefootmark{(a)} &    $<15$ &   -- & \citet{2021AJ....161..248W}\\
         51354 & \object{QY~Gem} &    $43.5\pm2.5$ &   --\tablefootmark{(a)} &    $102\pm4$ &   -- & \citet{2021AJ....161..248W}\\
         55606 & \object{MWC~522} &    $40.9\pm2.5$ &   --\tablefootmark{(a)} &    $<24$ &   93.8\tablefootmark{b} & \citet{2021AJ....161..248W}\\
         58978 & \object{FY~CMa} & $45\pm5$ & $4.3\pm0.6$ & $41\pm5$ & 37.3 & \citet{2008ApJ...686.1280P} \\%P=37.3d
         60855 & \object{V378~Pup} &    $42.0\pm2.5$ &   --\tablefootmark{(a)} &    $<27$ &   $346$ & \citet{2021AJ....161..248W}\\
         113120 & \object{LS~Mus} &   $45.0\pm2.5$ &   --\tablefootmark{(a)} &   $<36$ &   -- & \citet{2021AJ....161..248W}\\
         137387 & \object{\textkappa~Aps} &   $40.0\pm2.5$ &   --\tablefootmark{(a)} &   $<17$ &   $84$ & \citet{2021AJ....161..248W}\\
         152478 & \object{V846~Ara} &   $42.0\pm2.5$ &    --\tablefootmark{(a)} &    $<15$ &   -- & \citet{2021AJ....161..248W}\\
         157042 & \object{\textiota~Ara} &    $33.8\pm2.5$ &   --\tablefootmark{(a)} &   $<36$ &   -- & \citet{2021AJ....161..248W}\\
         157832 & \object{V750~Ara} & -- & -- & -- & -- & \citet{2018ApJ...853..156W}\\
         191610 & \object{28~Cyg} & -- & -- & -- & -- & \citet{2018ApJ...853..156W}\\
         194335 & \object{V2119~Cyg} &    $43.5\pm2.5$ &   --\tablefootmark{(a)} &   $<15$ &   $60.3$ & \citet{2021AJ....161..248W}\\
         200120 & \object{59~Cyg} & $52.1\pm4.8$ & $5.0\pm1.0$ & $<40$ & 28.2 & \citet{2013ApJ...765....2P} \\%P=28.2d
         200310 & \object{60~Cyg} & $42\pm4$ & $>4.75$ & -- & 147 & \citet{2017ApJ...843...60W} \\%P=147
         %214168 & \object{8~Lac~A} & -- & -- & -- & 15200\tablefootmark{(c)} & \citet{2017ApJ...843...60W} \\
         \hline
         %167128 & \object{HR~6819} & $16\pm1$ & $2.8\pm0.2$ & $<25$ & $40.335\pm0.007$ & \citet{bodensteiner}\\
         % &  & $16\pm1$ & $2.75\pm0.35$ & $<20$ & $40.3\pm0.3$ & \citet{elbadry2}\\
         %\hline
    \end{tabular}
    \tablefoot{\tablefoottext{a}{$4.75$ was assumed by \citet{2021AJ....161..248W}} \tablefoottext{b}{\citet{2018ApJ...865...76C}}} %\tablefoottext{c}{for a possible outer component\citet{2019AJ....158..167T}}}
    \label{tab:phiPer}
\end{table*}

\begin{figure*}
    \centering
    \includegraphics[width=\hsize]{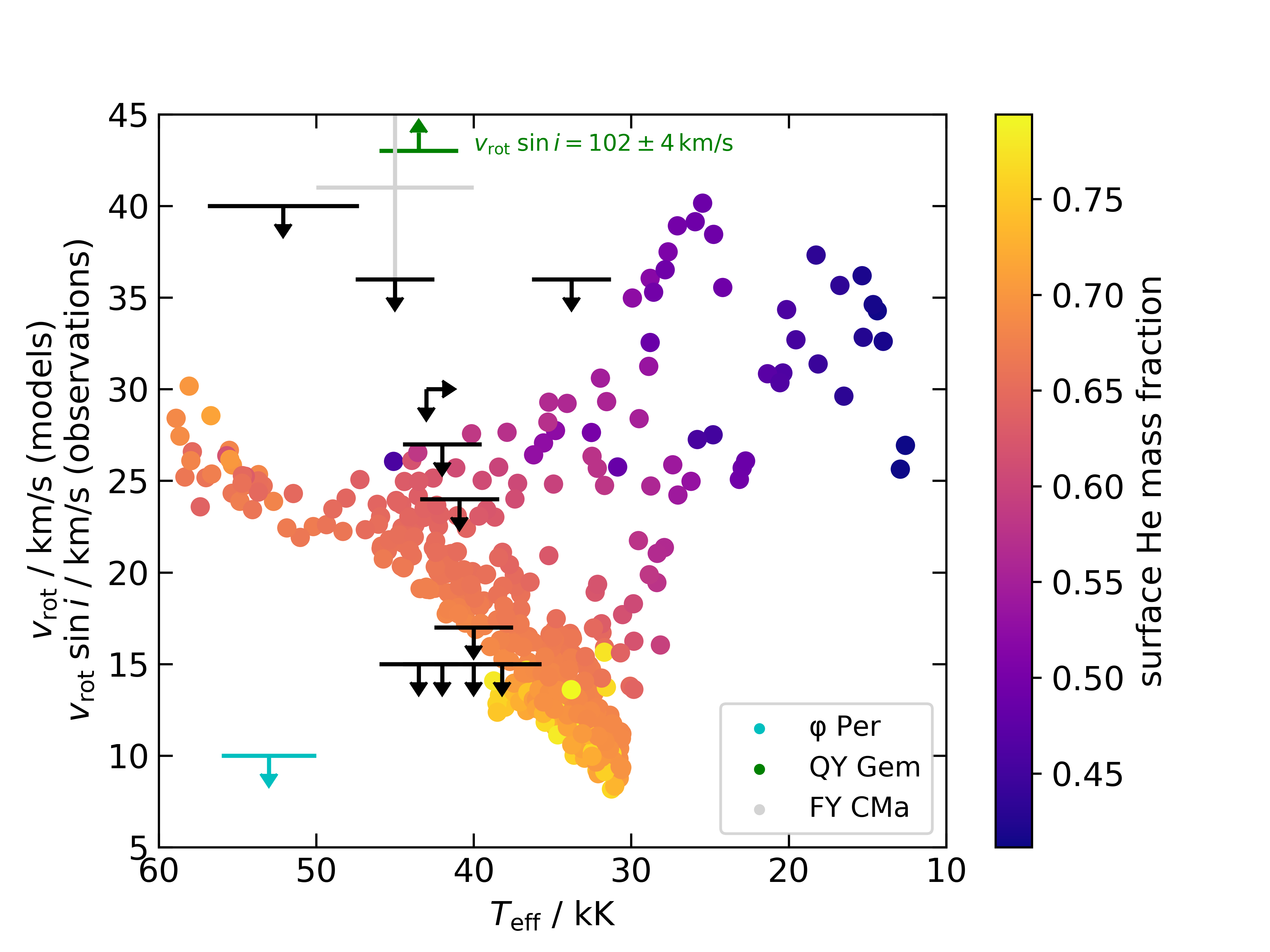}
    \caption{Effective temperature, rotational velocity, and surface helium abundance of our stripped core helium burning models at time of minimum radius (only systems with an initial donor mass below $10\ms$ are shown). The values measured for the sdOB~stars in the known sdOB+Be stars (references in Table~\ref{tab:phiPer}) are also shown. The rotational velocity of QY~Gem ($102\pm4\kms$, green) lies beyond the figure's limit.}
    \label{fig:sdOB}
\end{figure*}

As the stripped stars of LB-1 and HR~6819 contract further, they will become sdOB~stars in a \textphi~Persei-like binary (i.e. a sdOB+Be-system). While this configuration is expected to be common for Be-stars \citep{elbadry2,2021AJ....161..248W}, it is hard to identify as the Be-star outshines the subdwarf, and so far fewer than 20 such systems are known (16 confirmed plus 3 candidates) and have been spectroscopically analysed \citep{2018ApJ...853..156W, 2018ApJ...865...76C, 2021AJ....161..248W}. We compiled the effective temperature, gravity, and rotation (if available) of the subdwarfs as well as the orbital period of the systems in Table~\ref{tab:phiPer}. For two sdOB~stars, the projected rotational velocity is known, but for 12 we know only an upper limit. We show them in Fig.~\ref{fig:sdOB} together with the calculated values from our binary models discussed in Sect.~\ref{sec:res-pro}. In this diagram we show the stripped star models at the time of smallest radius during core helium burning (see  Fig.\,2). This relatively long-lasting phase is close to the helium main sequence.

Figure~\ref{fig:sdOB} shows that the helium-rich models follow a rather narrow correlation, with the hotter models rotating faster. Models with a lower surface helium abundance deviate from this pattern as their larger envelopes make them larger, and thus the surface temperature is cooler. The values of the observed sdOB~stars fall well within the predicted range. However, they do not follow the trend drawn by the models. Since for most of the stars only an upper limit of the projected rotation velocity is known, a more detailed comparison is difficult.

Three stars are remarkable in Fig.~\ref{fig:sdOB}: \textphi~Per, QY~Gem, and FY~CMa. \textphi~Per shows a significantly slower projected rotation than expected. While the inclination of the system could be small, it could also imply that its stripped star is past core helium exhaustion (see Fig.~\ref{fig:Tv}). Disregarding rotation, \citet{abel} found that a helium shell-burning model agrees with \textphi~Per. However, Fig.~\ref{fig:Tv} implies that a rotation velocity below 10\,km/s is only achieved during the fast final contraction towards the WD stage. QY~Gem and FY~CMa have a relatively high rotation velocity, which is not explained by our models. Their sdOB star could be spun up by accretion from the Be star's disc, as proposed by \citet{2021AJ....161..248W}.

\subsection{Angular momentum transport in stellar models} \label{sec:dis-calc}
In the course of stellar evolution, the core of a star  contracts, while the envelope, separated from the core by a jump in chemical composition,  expands. This happens already during the main-sequence evolution \citep{2020A&A...633A.165H}, and much more so after core hydrogen exhaustion, after which the hydrogen burning shell source strongly adds to the entropy jump between core and envelope. In rotating stars, one may therefore expect the core rotation frequency to increase with time, while the envelope does the opposite.

During the last decades, observational evidence has accumulated showing that this is not the case, or at least to a much lower degree than   expected from local angular momentum conservation. The specific angular momentum in upper main-sequence stars is typically $10^{17}...10^{18}\,\mathrm{cm^2/s}$, while it is three to four orders of magnitude smaller in WDs \citep{2008A&A...481L..87S} and young neutron stars \citep{heger2005}. This shows that angular momentum is drained from the stellar cores, and transported into the stellar envelopes during their evolution. In red giant stars this process has been traced as a function of the evolutionary stage through the analyses of their oscillations \citep{2012A&A...548A..10M,2014A&A...564A..27D,2018A&A...616A..24G}. The physical mechanism responsible for this angular momentum transport is still debated, with magnetic torques and internal waves being the strongest candidates \citep{2019ARA&A..57...35A}. Since the latter still have to be explored systematically in stellar evolution calculations, we focus here on models employing magnetic torques.

Several groups have studied the angular momentum transport imposed by the  magnetic torques as proposed by \citet{spruit}. \citet{2004A&A...422..225M} demonstrated that its inclusion in single-star calculations leads to near solid body rotation during the main-sequence evolution. In contrast, in models that include only hydrodynamic angular momentum transport \citep{heger2000}, the  core and envelope  rotate nearly rigidly, but each with its individual rotation frequency, and the difference amplifies during the  main-sequence evolution. Here the non-magnetic transport mechanisms are not able to overcome the gradient of entropy and mean molecular weight that separates core and envelope. After central hydrogen exhaustion the difference in rotational frequency grows to several orders of magnitude. Our models with and without the magnetic transport follow these patterns.

\citet{heger2005} showed that the magnetic torques proposed by \citet{spruit} remove angular momentum from the core of main-sequence models compared to non-magnetic models \citep[see also][]{2006A&A...460..199Y}. Most of this happens between core hydrogen depletion and helium ignition. As we observe in our models, \citet{heger2005} noted that the helium burning core rotates nearly rigidly. Similarly, \citet{2008A&A...481L..87S} demonstrated that calculations incorporating magnetic torques lead to WD spins close to the observed values, while WD models resulting from non-magnetic models are rotating orders of magnitude too rapidly. Our binary and single-star model behave in a  comparable manner.

The first binary calculations with magnetic transport were performed by \citet{2005A&A...435..247P}. They noted that the extraction of angular momentum from the cores renders the formation of long-duration gamma-ray bursts through this channel unlikely. As in single stars, magnetic torques during the main-sequence and early hydrogen shell-burning evolution are able to remove angular momentum from the stellar cores. Additionally, the authors find that the mass donors have an extremely slow rotation after RLO. \citet{2007A&A...465L..29C} calculated, as we do, a Case~B binary including magnetic transport, but for higher masses. Nevertheless, they found the donors to rotate slowly after RLO (see their Table~1). Both works are in agreement with our models.

\citet{2010ApJ...725..940Y} computed binary evolution models of type Ib/c supernova progenitors with and without magnetic angular momentum transport. Their magnetic models show that the cores lose, as in our models, large amounts of angular momentum during RLO. In Case~A models tides play a role; instead,  in Case~AB \citep[mass transfer subsequent to Case~A after central hydrogen depletion,][]{1967ZA.....65..251K} and B mass transfer,  magnetic transport is responsible for this angular momentum loss. This becomes evident through their non-magnetic model, where the core's angular momentum barely changes during Case~AB RLO. More recently, \citet{2020A&A...640L..18M} showed that magnetic angular momentum transport has an impact on the upper black hole mass-gap between $45\ms$ and $120\ms$.

As for massive stars, magnetic transport due to the  Spruit--Tayler dynamo has been extensively tested in low mass models at various stages,  on the main sequence and for the Sun \citep{2005A&A...440L...9E,2019A&A...626L...1E}, on the subgiant and red-giant branches \citep{2014ApJ...793..123M}, and for WDs \citep{2008A&A...481L..87S,2017A&A...602A..55N}. For more evolved stars the results become more complex. \citet{2014ApJ...788...93C} demonstrated that the original Spruit--Tayler dynamo alone is not able to reproduce the observed angular momentum loss of the core of red giants and helium core-burning low mass stars. Other studies \citep{2013A&A...555A..54C, 2015A&A...579A..31B, 2015ApJ...799...85W, 2016A&A...589A..23S, 2017A&A...599A..18E, 2019A&A...626A.121O} enforce the conclusion that additional angular momentum transport may be required. \citet{2019MNRAS.485.3661F} have reanalysed the formulation of the Spruit--Tayler dynamo and proposed a revision, which, as they demonstrate in stellar evolution calculations, allows  the observed core angular momentum evolution in red giants to be reproduced.

While some authors debate the functionality of the Spruit--Tayler dynamo on theoretical grounds \citep{2006A&A...449..451B,2007A&A...474..145Z} or question its existence based on observations \citep[e.g.][]{2010ApJ...716.1269D}, angular momentum transport by magnetic torques from toroidal B-fields are undisputed. \citet{2021A&A...646A..19T} presented magneto-rotational stellar evolution calculations in which the internal magnetic field evolution is described by two time-dependent differential equations, which are solved along with the stellar structure equations. Their models obtain angular momentum transport by magnetic torques on the Alfv\'en timescale, which was shown to be able to reproduce the red giant observations. Overall, while the description of  angular momentum transport by magnetic fields in 1D stellar evolution models is still improving, this mechanism is a strong candidate to provide a realistic description of the evolution of the internal rotation of evolved stars.

\subsection{Orbital evolution and mass transfer efficiency}\label{sec:dis-evo}

\begin{figure}
    \includegraphics[width=1.1\hsize]{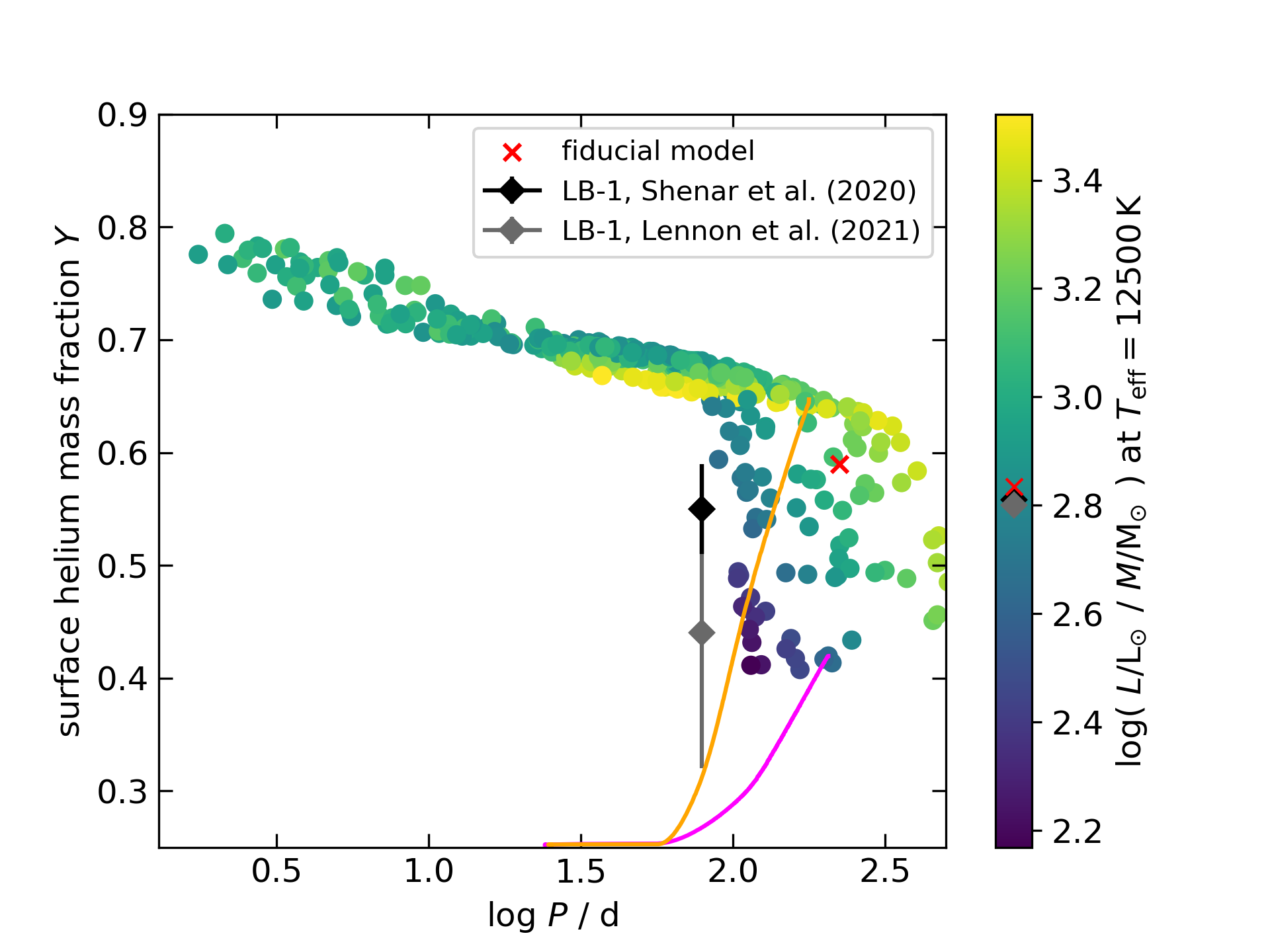}
    \caption{Period, surface helium abundance, and luminosity-to-mass ratio during the contraction phase after RLO, when the stripped star has obtained an effective temperature of 12\,500\,K (close to that derived for LB-1). The black and grey symbols indicate the values for LB-1, as derived by \citet{shenar} and \citet[][almost superimposed in the colour bar]{lennon}. The fiducial model is shown in red. The orange curve (initially $4.27\ms$ + $4.07\ms$ at $P_\mathrm{orb} = 24.2\days$) shows the surface helium evolution of a mass donor, which does not develop convection in its outer layers during RLO. In the magenta curve (initially $5.86\ms$ + $4.20\ms$ at $P_\mathrm{orb} = 25.4\days$) convection sets in.}
    \label{fig:PeHe}
\end{figure}

In Sects.~\ref{sec:res-pro} and \ref{sec:res-rot2} we presented our fiducial binary evolution model, which reproduces the observed properties of the stripped star in LB-1. However the final orbital period we found did not agree with the observed value. In Fig.~\ref{fig:PeHe} we show the orbital period, helium abundance, and luminosity-to-mass ratios for all the models discussed in Sect.~\ref{sec:res-pro}, at a time when the mass donor after RLO has reached $T_\mathrm{eff} = 12\,500\,\mathrm{K}$. We find no model in our set with all \emph{three} observed properties of the stripped star fitting to LB-1, although some come close to it. 

In Fig.~\ref{fig:PeHe} we can identify two groups, one group above $Y=0.6$ with a tight correlation between orbital period and helium abundance, and a more scattered group at periods above 80\,d. The reason for this scatter is the development of convection in the mass donors' outer layers during RLO. Donors in close orbits have small Roche lobes, and  therefore they cannot expand enough to become cool enough to develop a notable envelope convection zone, while in wide orbit binaries a convective envelope can develop  that mixes hydrogen into the regions that  were helium-enriched during the main-sequence evolution. After the mass transfer, these regions are exposed at the surface, resulting in a lower surface helium abundance compared to models for which convection does not occur.

This can be seen from the two curves indicated in Fig.~\ref{fig:PeHe}. They show the evolution of the surface helium abundance during the RLO for two selected models. As the donor loses mass, the orbit widens and at some point helium-enriched layers appear at the surface. In the orange curve in Fig.~\ref{fig:PeHe} the convection zone never extends into the layers, which were  helium-enriched during hydrogen burning, and thus the helium abundance grows quickly as deeper regions are uncovered. In the magenta curve convection sets in in the outer layers before the helium-rich layers are exposed. The convective region grows and eventually its lower boundary reaches the helium-enriched region, and thus hydrogen-rich material is mixed into that region. If this region is exposed later, its helium abundance does not reach  values as high as in the case without convection. Thus, the magenta curve is not as steep as the orange one, and models with convection do not reach final helium values as high as those without convection, and the correlation between the  orbital period and helium abundance disappears.

In Fig.~\ref{fig:PeHe} we also plot the measurements of \citet{shenar} and \citet{lennon}. Both lie slightly below the low period interval covered by the scattered group. The value of \citet{shenar} fits well to the helium abundance and the luminosity-to-mass ratio of the fiducial model, but it does not reproduce the orbital period. The helium abundance measured by \citet{lennon} deviates more from the model, but has an uncertaintly that is quite  large. In Fig.~\ref{fig:PeHe} their data point lies closest to models with luminosity-to-mass ratios certainly below their measurement. If  all the models were shifted by about 0.5~dex to the left, the fiducial model's orbital period would be consistent with the observations and the two observations   surrounded by models with luminosity-to-mass ratios in agreement with them.

We attribute the mismatch in orbital period to the uncertain mass and angular momentum loss of the binary during the RLO. Our fiducial model for example increases its orbital period from an initial value of  $16\days$ to $223\days$, while the orbital period of LB-1 is $79\days$ (Table~\ref{tab:atmos}). A prescription including either a similar mass loss from the binary which carries more angular momentum per mass unit or a lower mass loss from the binary systems would lead to models with shorter periods and in better agreement to the observations.  \citet{abel} and \citet{2007ASPC..367..387P} for \textphi~Per, as well as \citet{2014ApJ...796...37S} for Be star binaries, report a preference for non-conservative mass transfer with higher mass transfer efficiencies than those in our highly non-conservative models ($<5\%$ mass transfer efficiency). Similarly, \citet{2021ApJ...908...67S} performed a population synthesis analysis of LB-1 assuming different mass transfer models and found a strong preference for non-conservative mass transfer. We decided against searching for a model with matching orbital period, as both the accretion efficiency and the specific angular momentum of the expelled matter are unknown. The large uncertainty of the mass measurement of the Be~star \citep{shenar,lennon} would not allow  the degeneracy impeding a sound result to be lifted.

%One caveat is that the donor's helium abundance after the RLO may dependent on the evolution of the Roche volume during RLO, which would change under a different mass transfer prescription, and hence a simple shift of the simulations along the horizontal axis in Fig.~\ref{fig:PeHe} may be too simple.

\citet{bodensteiner} and \citet{elbadry2} provide MESA models for HR~6819. \citet{bodensteiner} restrict themselves to an initial mass ratio of 1/3, and find that initially very close systems, which evolve through Case~A mass transfer, can reproduce the observations. In our simulations, Case~A systems show  orbital periods that are too short to be relevant for LB-1 and HR~6819. However, the helium abundance in the stripped star model ($Y=0.87$) of \citet{bodensteiner} does not match the observations of \citet[][$Y=0.54\pm0.11$, Table~\ref{tab:atmos_hr}]{elbadry2}. \citet{elbadry2} provide a set of calculations with varying mass transfer efficiency, some of which lead to solutions comparable with LB-1 and HR~6819. However, in their analysis, and  in that of \citet{eldridge}, the helium abundance is not considered. We demonstrated that this quantity is very constraining, as it strongly correlated with the post-RLO orbital period and with the rotational velocity in most of the model parameter space. Our results imply that it may be feasible to compare binary models with varying mass transfer efficiencies to measurements of period and surface helium abundance for LB-1, HR~6819, and the known sdOB+Be systems in order to constrain the physical mechanisms driving mass transfer.

\section{Conclusions}\label{sec:concl}

In this study our aim was to constrain the angular momentum transport mechanisms in the stellar interior by modelling the mass donor of the binary system LB-1, composed of a stripped B-type star and a Be~star. To this end, we analysed a large grid of MESA binary evolution models to investigate the rotational evolution of the mass donors after mass transfer, and to identify models corresponding to the evolutionary phase of LB-1 and similar binaries in the stripped star scenario. We focused on the luminosity-to-mass ratio of the stripped star which is observationally well determined. We found that in Case~B models (mass transfer after donor core hydrogen exhaustion) this parameter is uniquely determined by the donor's initial mass at the moment of Roche-lobe detachment. However, the ensuing drop in the luminosity-to-mass ratio also depends  on the mass of the remaining envelope, which is indicated by the final surface helium abundance, which is set in turn by the initial orbital period. Based on the observed luminosity-to-mass ratio and surface helium abundance, we obtained an initial mass for the stripped star in LB-1 of about $4\ms$. 

We examined the internal rotation, and the evolution of the surface rotation rate of our donor star models. To do so we calculated MESA models including magnetic angular momentum transport by the Spruit--Tayler dynamo, which removes angular momentum from the stellar core during and after central hydrogen exhaustion and yields low surface rotational velocities in the stripped mass donor after RLO. The braking, which is caused by tidal forces, leads to about the same core angular momentum and final WD spin as obtained in single-star models where the core rotation is braked by the expanded envelope and the mass growth of the core. In the binary case the envelope accelerates again while the star contracts towards sdOB~phase. Our results agree qualitatively with observations, suggesting that angular momentum is removed from the stellar cores  by magnetic angular momentum transport through the Spruit--Tayler dynamo until well into the hydrogen shell-burning stage, based on asteroseismic results and spin rates of WDs and neutron stars. 

We considered models with and without angular momentum transport by magnetic torques, and models with different initial rotation rates. When comparing the rotation velocity of our models during the contraction phase to the observed value for LB-1 \citep{shenar,lennon}, we found that it can be reproduced only by our magnetic models, independent of the initial rotation. Our models predict a relation between the orbital period, temperature, and equatorial rotation of contracting post-RLO stars, which may be used to determine the mass transfer efficiency during RLO. A comparison to a larger sample of observed sdOB+Be-systems, where often only upper limits for the rotational velocities of the stripped star are known, shows a broad agreement with our models. 

Furthermore, we find evidence that our employed mass transfer scheme   underestimates the mass transfer efficiency during RLO, such that our models can only marginally reproduce the observed orbital period of LB-1 and similar systems. This suggests, in line with previous studies, that the Be~stars in the considered systems have accreted notable amounts of material. A population synthesis study should be able to determine the accretion efficiency of the RLO and the specific angular momentum the expelled matter carries. The surface helium abundance of stripped stars, together with their orbital periods, offers a new tool to tightly constrain the accretion efficiency in mass transferring binaries.

\begin{acknowledgements}
 We would like to thank Tomer Shenar and Julia Bodensteiner for discussing the observational aspects with us as well as Pablo Marchant for sharing his MESA framework with us.
\end{acknowledgements}

\bibliographystyle{aa}
\bibliography{LB1}

\newpage
\onecolumn
\appendix
\section{Variations in initial rotation and dynamo}

\begin{figure*}[h]
    \includegraphics[width=0.5\hsize]{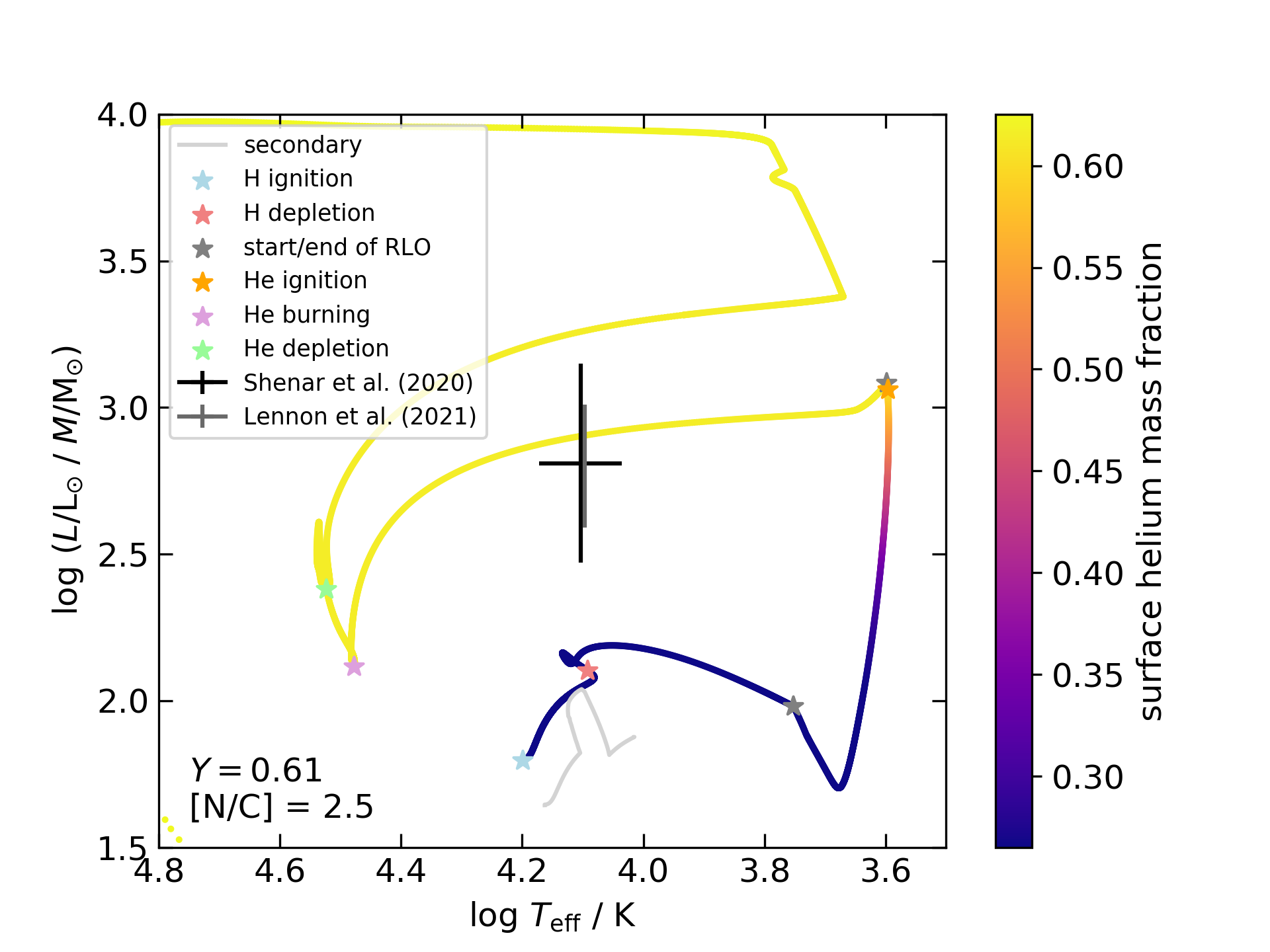}
    \includegraphics[width=0.5\hsize]{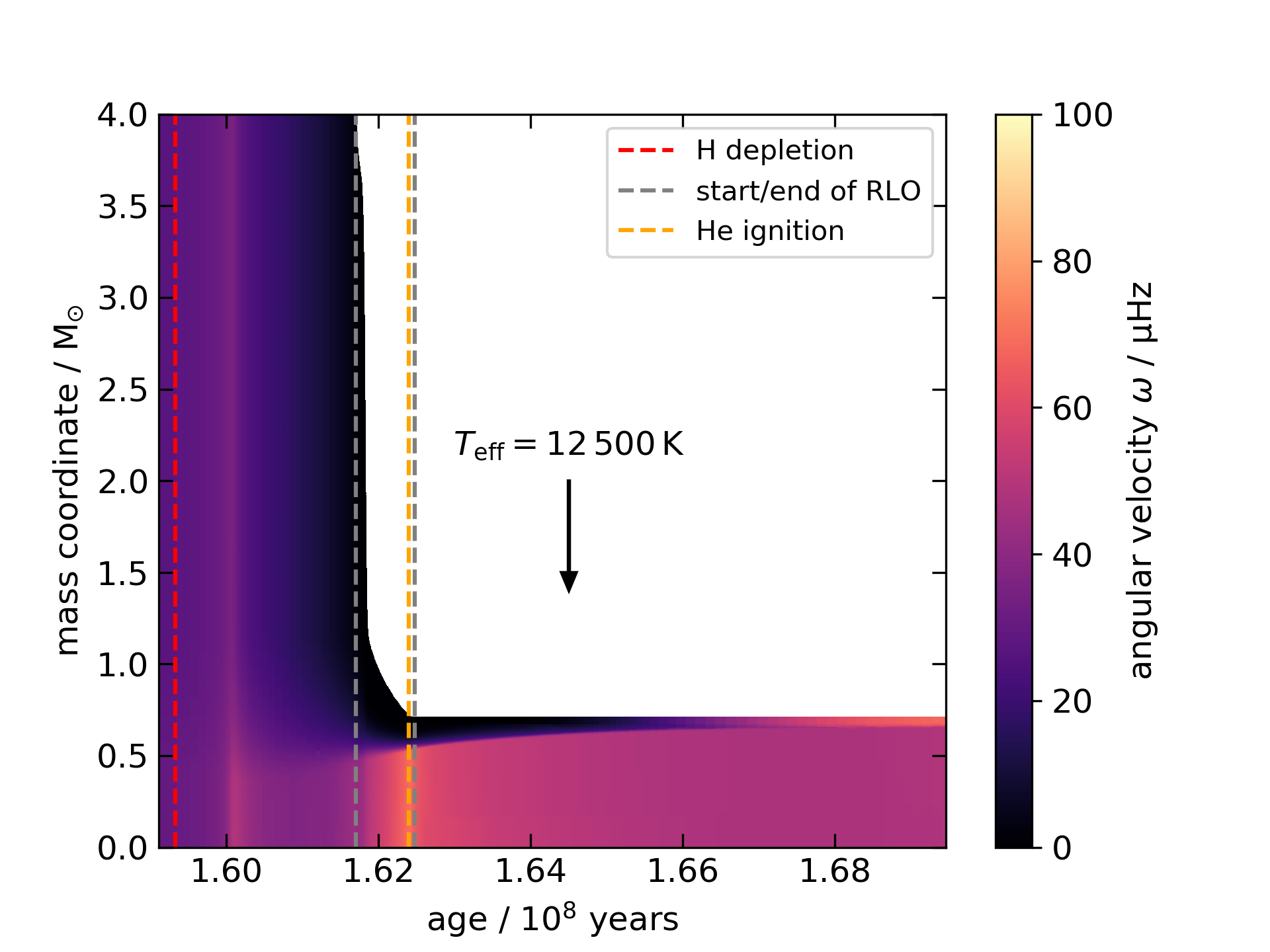}
    \includegraphics[width=0.5\hsize]{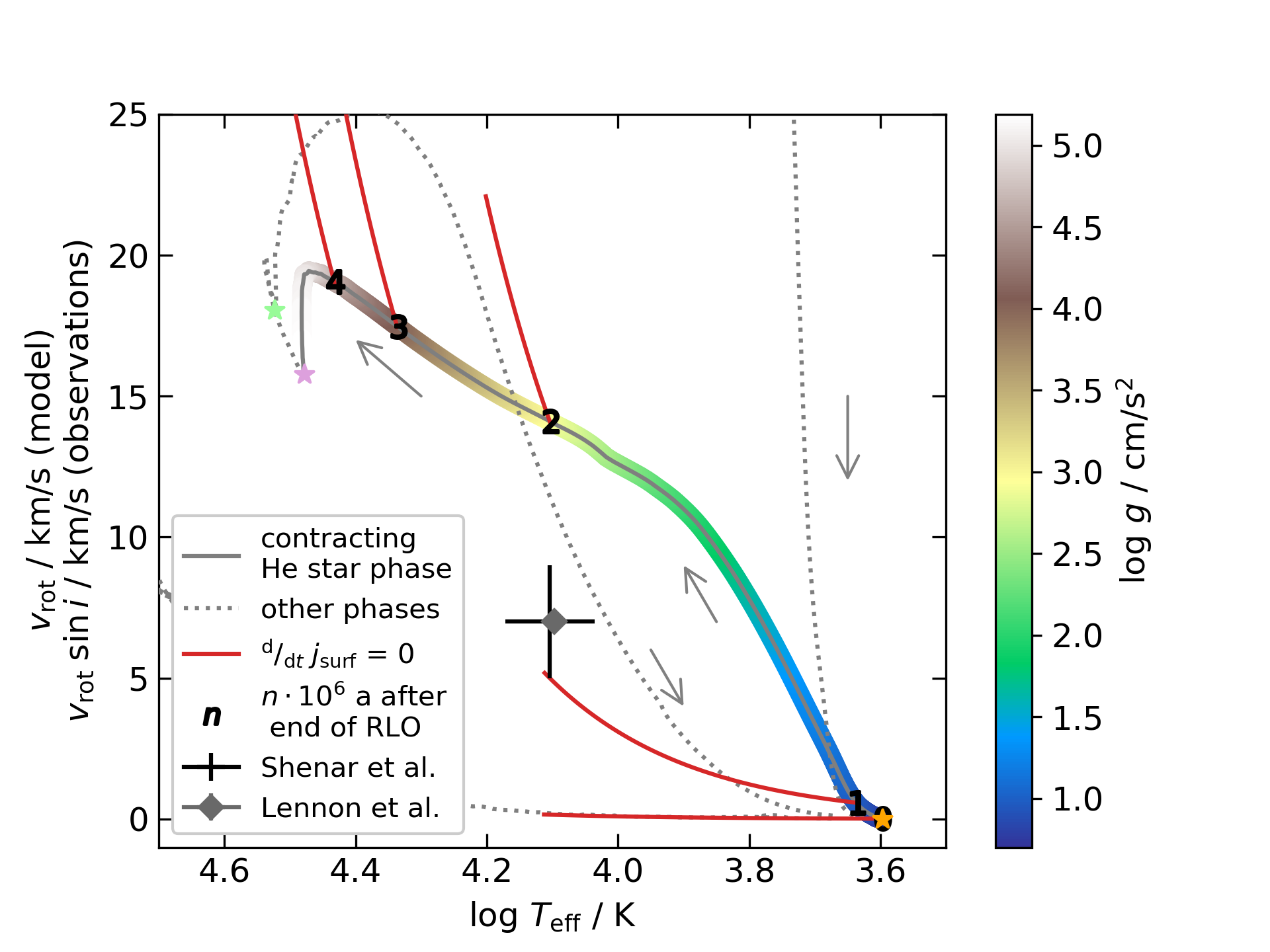}
    \includegraphics[width=0.5\hsize]{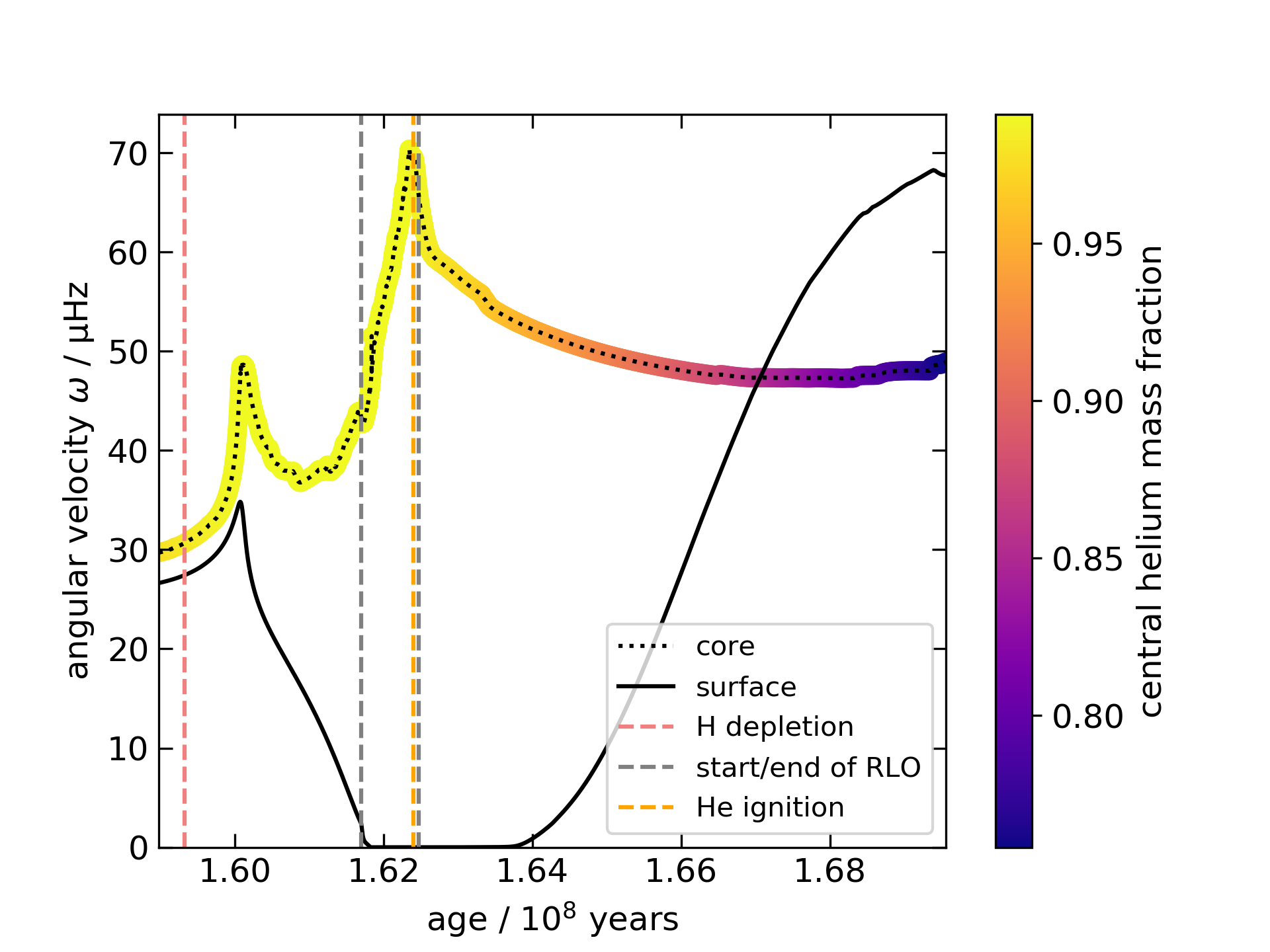}
    \caption{Same as Figs.~\ref{fig:hrd}, \ref{fig:kippen}, and \ref{fig:Tv}, but with an initial rotation of 20\% of the critical rotation.}
    \label{fig:20vvc}
\end{figure*}

\begin{figure*}[h]
    \includegraphics[width=0.5\hsize]{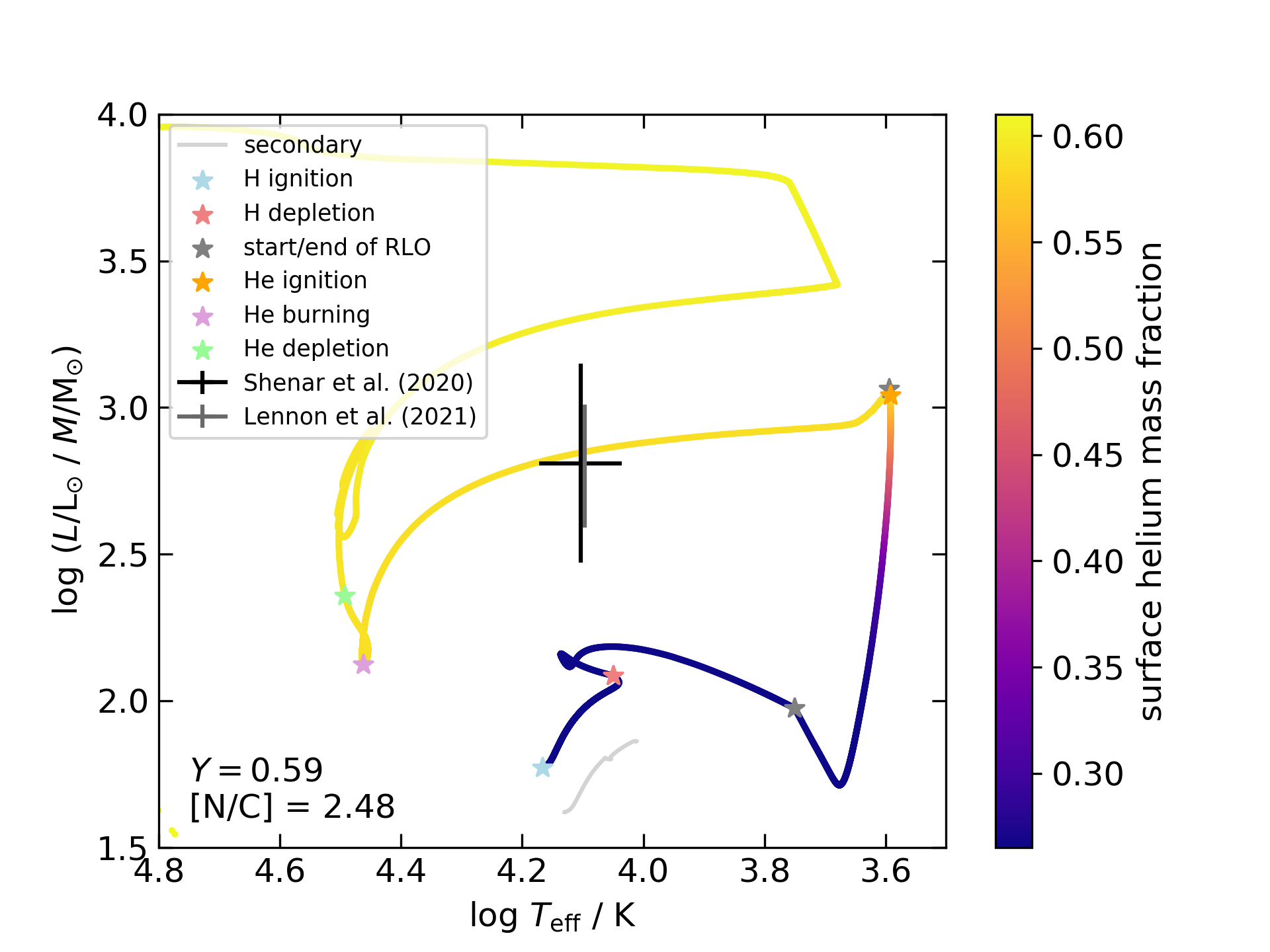}
    \includegraphics[width=0.5\hsize]{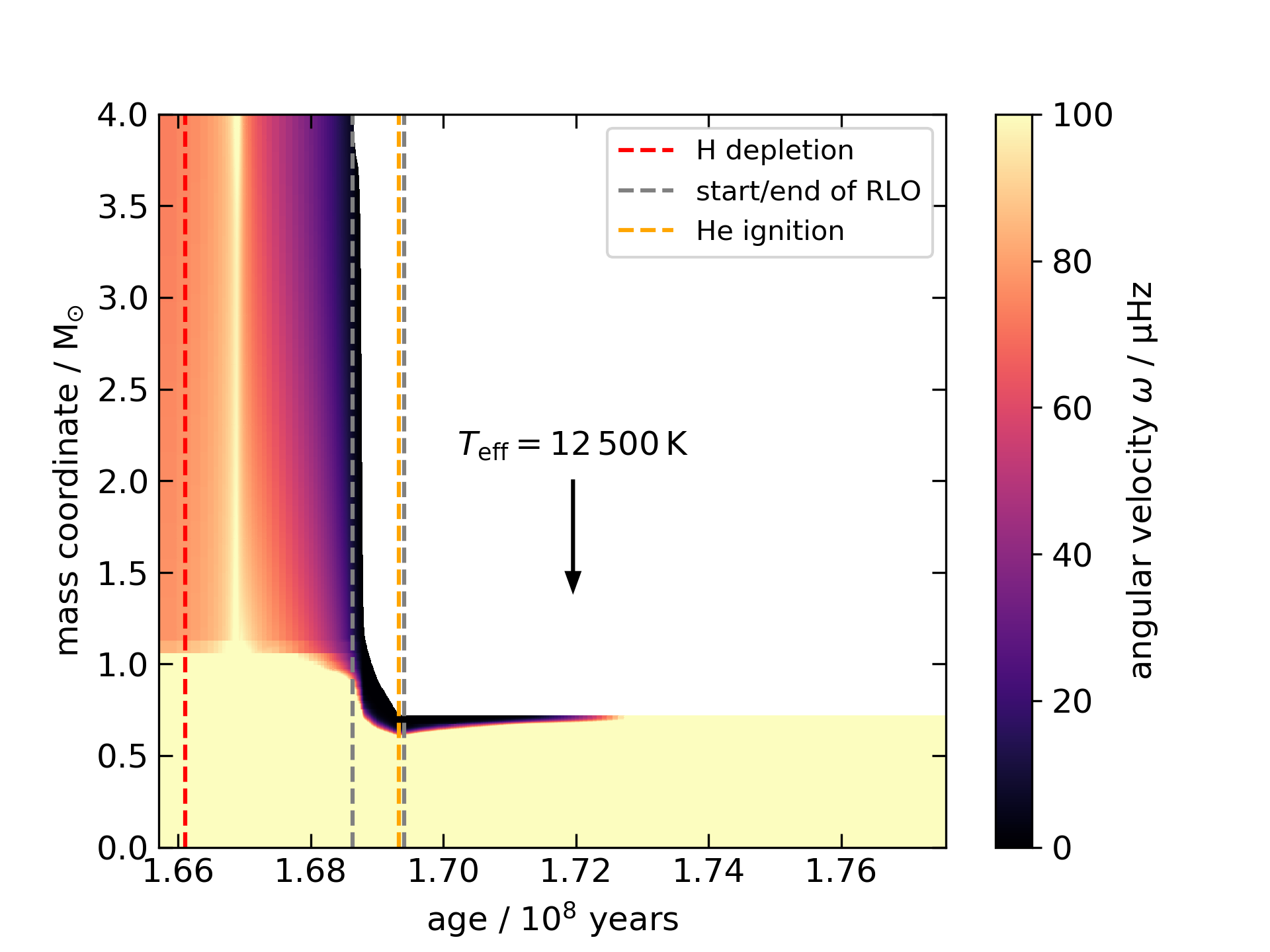}
    \includegraphics[width=0.5\hsize]{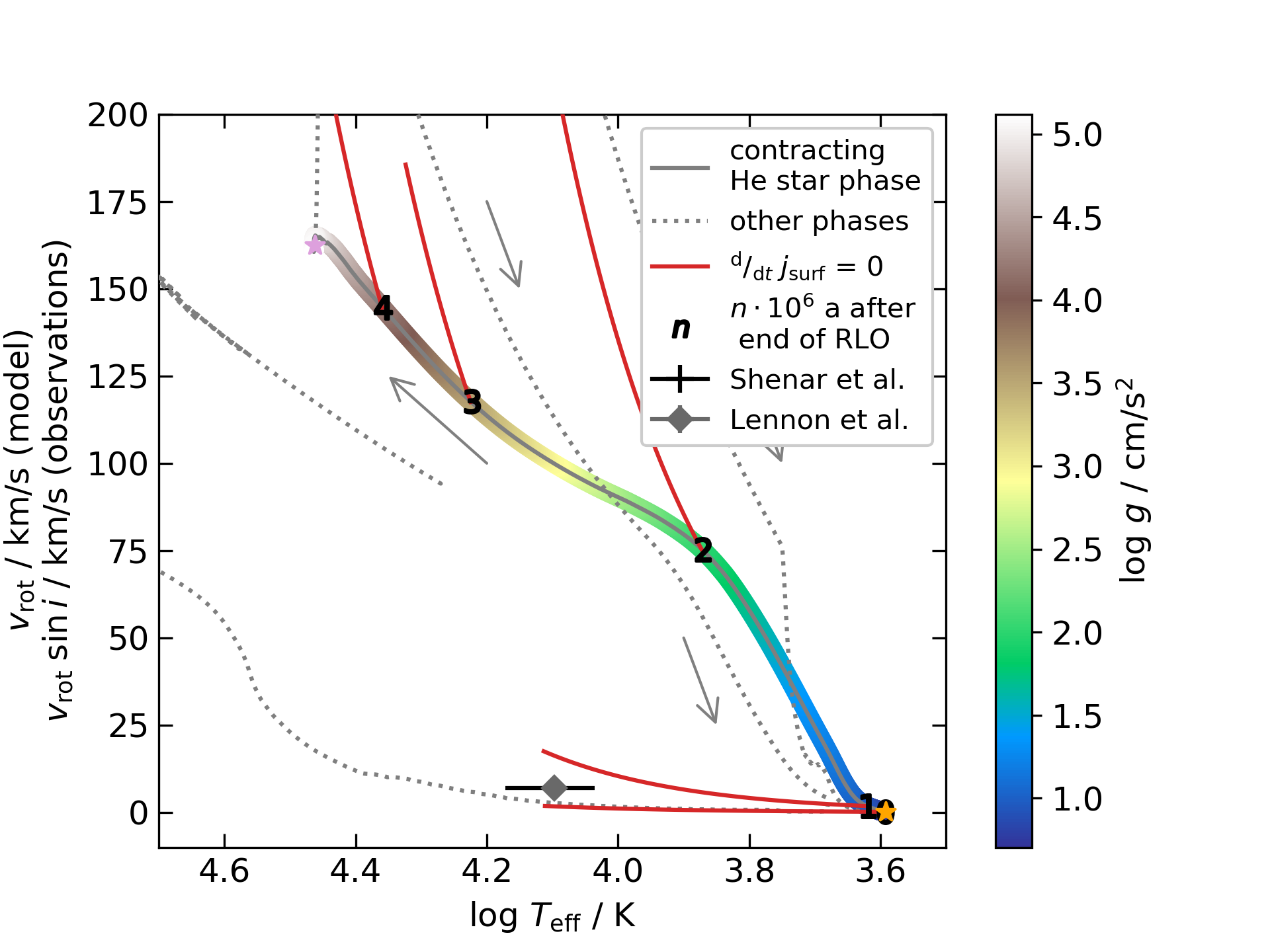}
    \includegraphics[width=0.5\hsize]{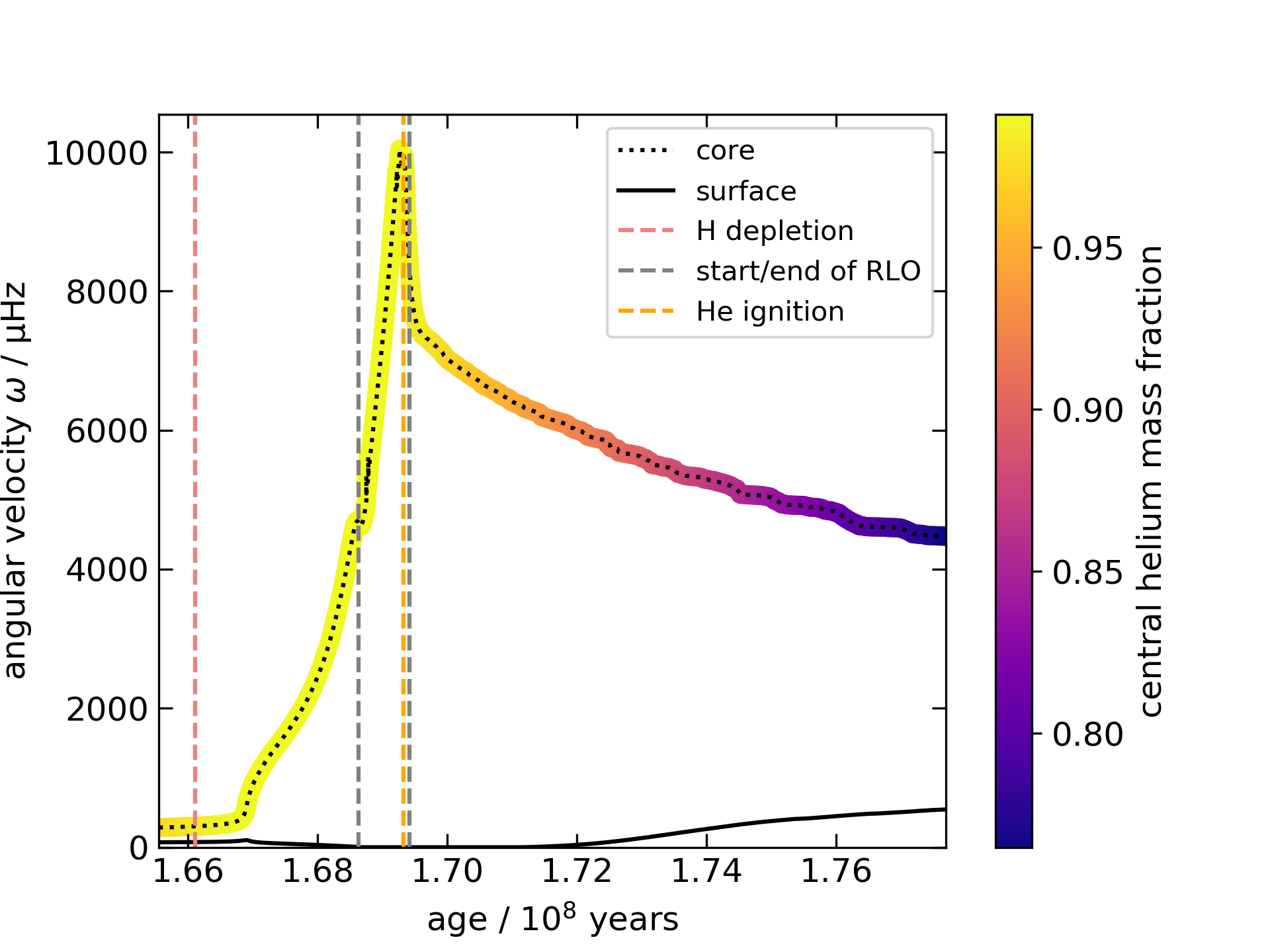}
    \caption{Same as Figs.~\ref{fig:hrd}, \ref{fig:kippen}, and \ref{fig:Tv}, but without the  Spruit--Tayler dynamo. The top right panel only depicts angular velocities up to $100\,\text{\textmu Hz}$.}
    \label{fig:noB}
\end{figure*}

\begin{figure*}[h]
    \includegraphics[width=0.5\hsize]{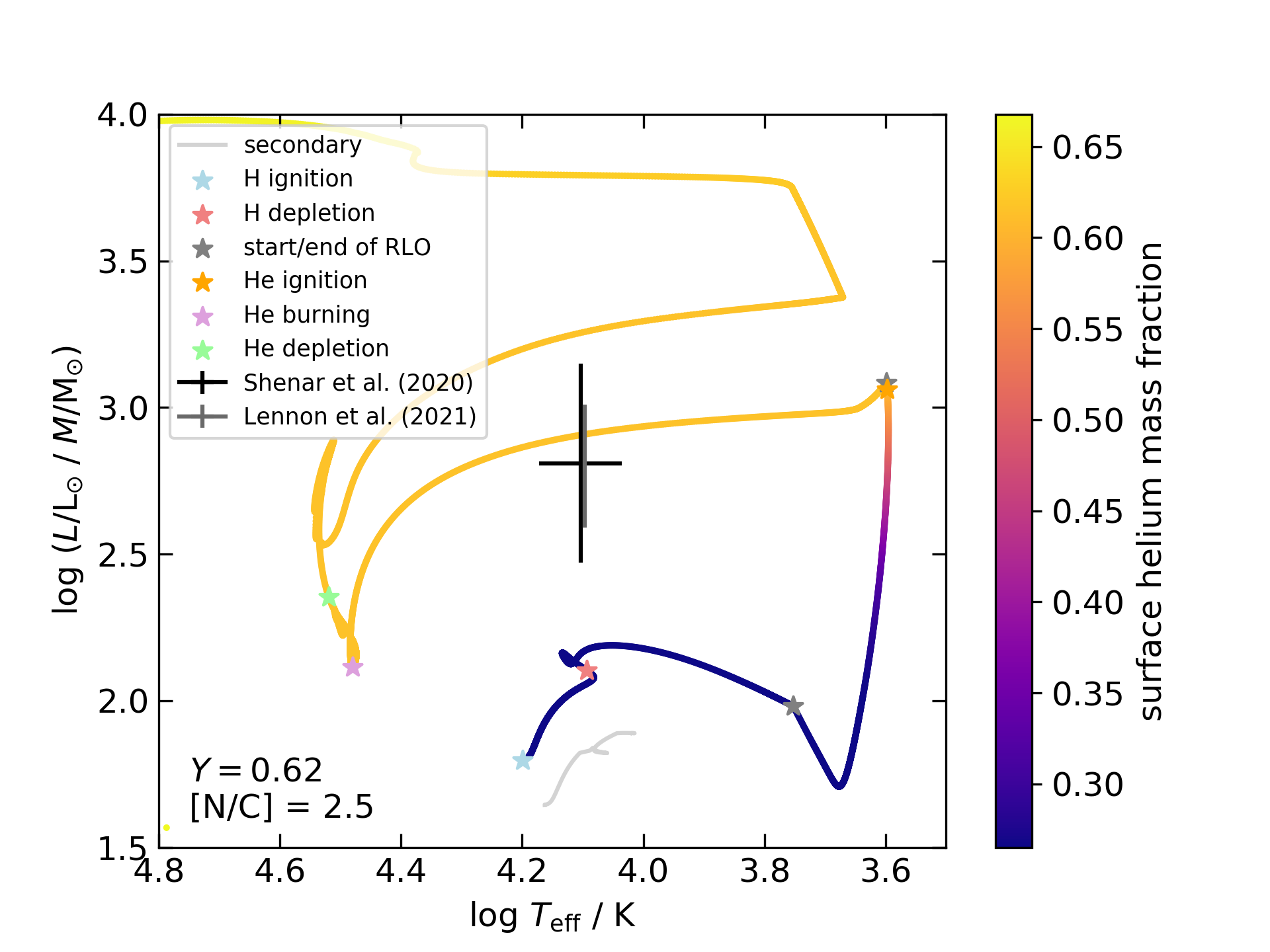}
    \includegraphics[width=0.5\hsize]{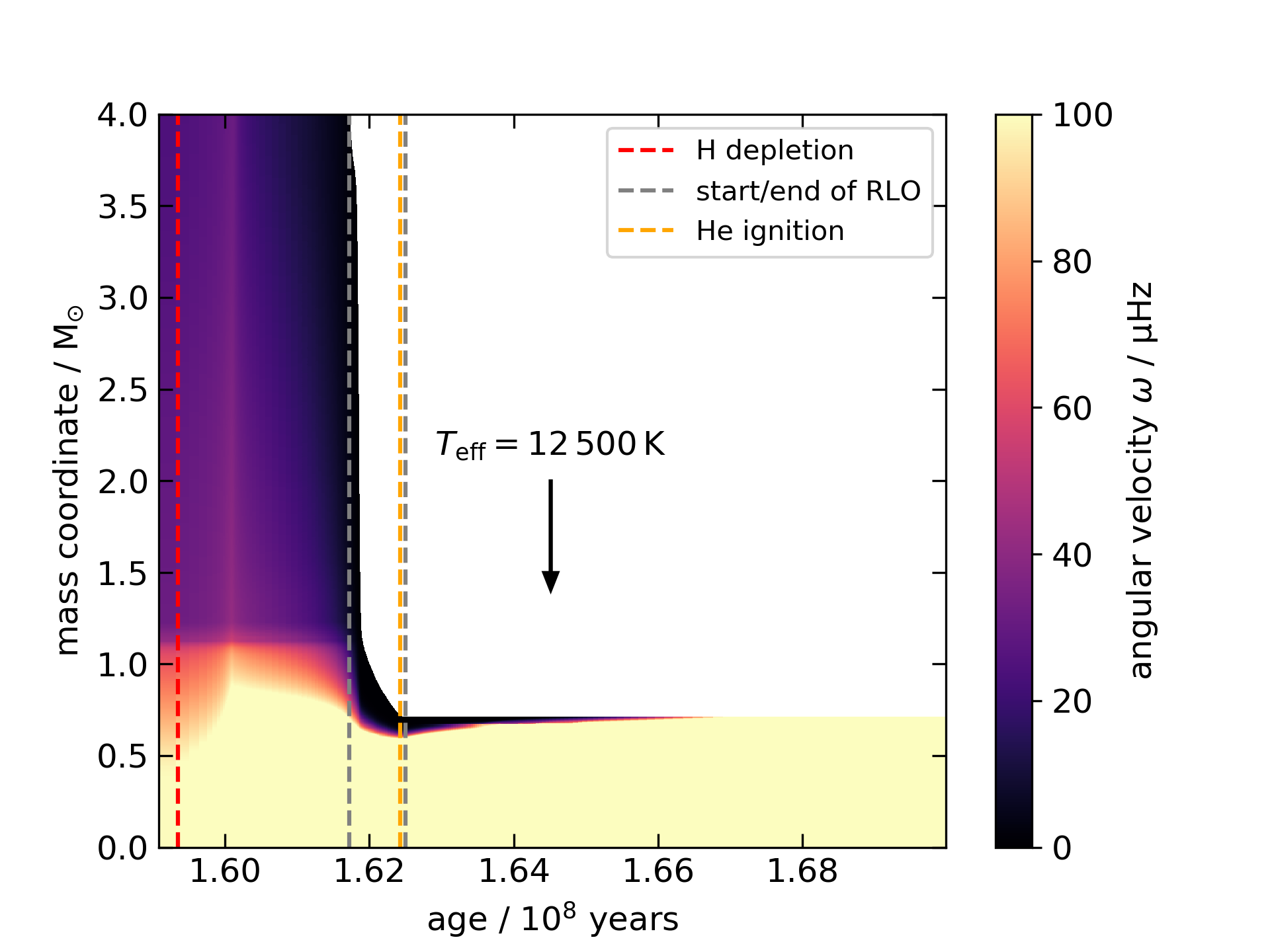}
    \includegraphics[width=0.5\hsize]{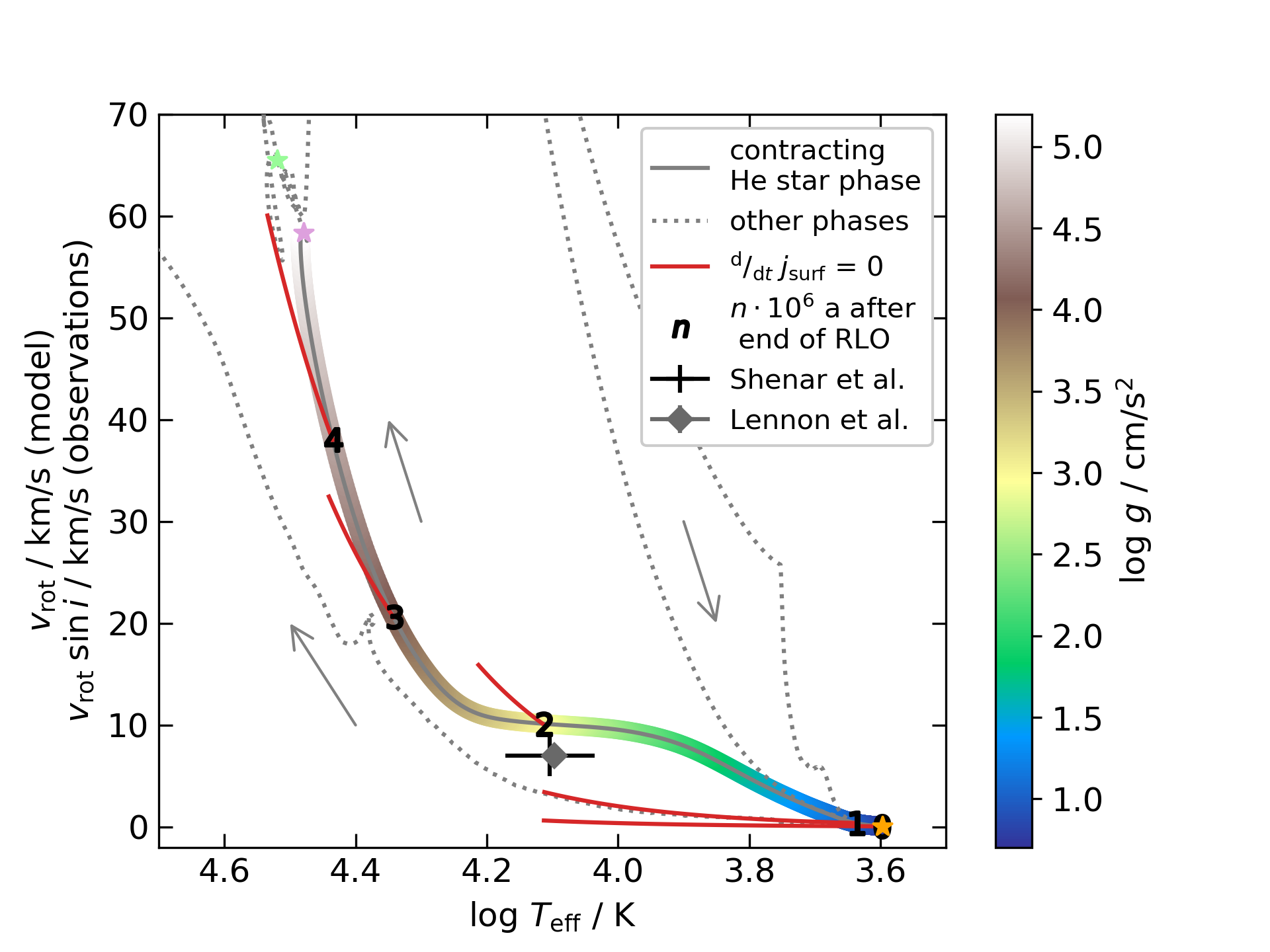}
    \includegraphics[width=0.5\hsize]{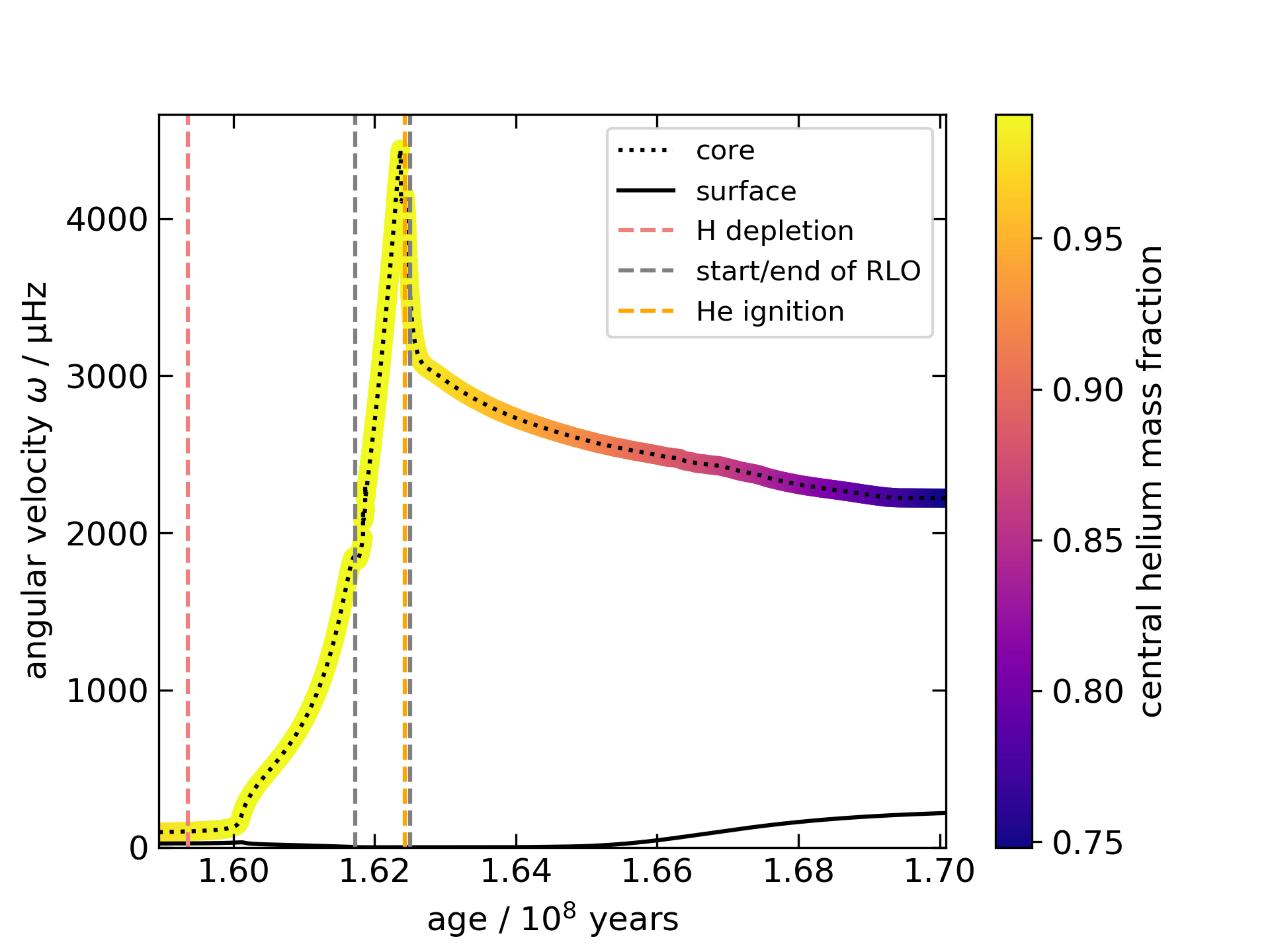}
    \caption{Same as Figs.~\ref{fig:hrd}, \ref{fig:kippen}, and \ref{fig:Tv}, but with an initial rotation of 20\% of the critical rotation and without the Spruit--Tayler dynamo. The top right panel only depicts angular velocities up to $100\,\text{\textmu Hz}$.}
    \label{fig:20noB}
\end{figure*}

\newpage \quad \newpage \quad \newpage
\section{Predictions for orbital period, rotational velocity, and surface helium abundance of donor star models contracting after RLO}

\begin{figure*}[h]
    \includegraphics[width=0.5\hsize]{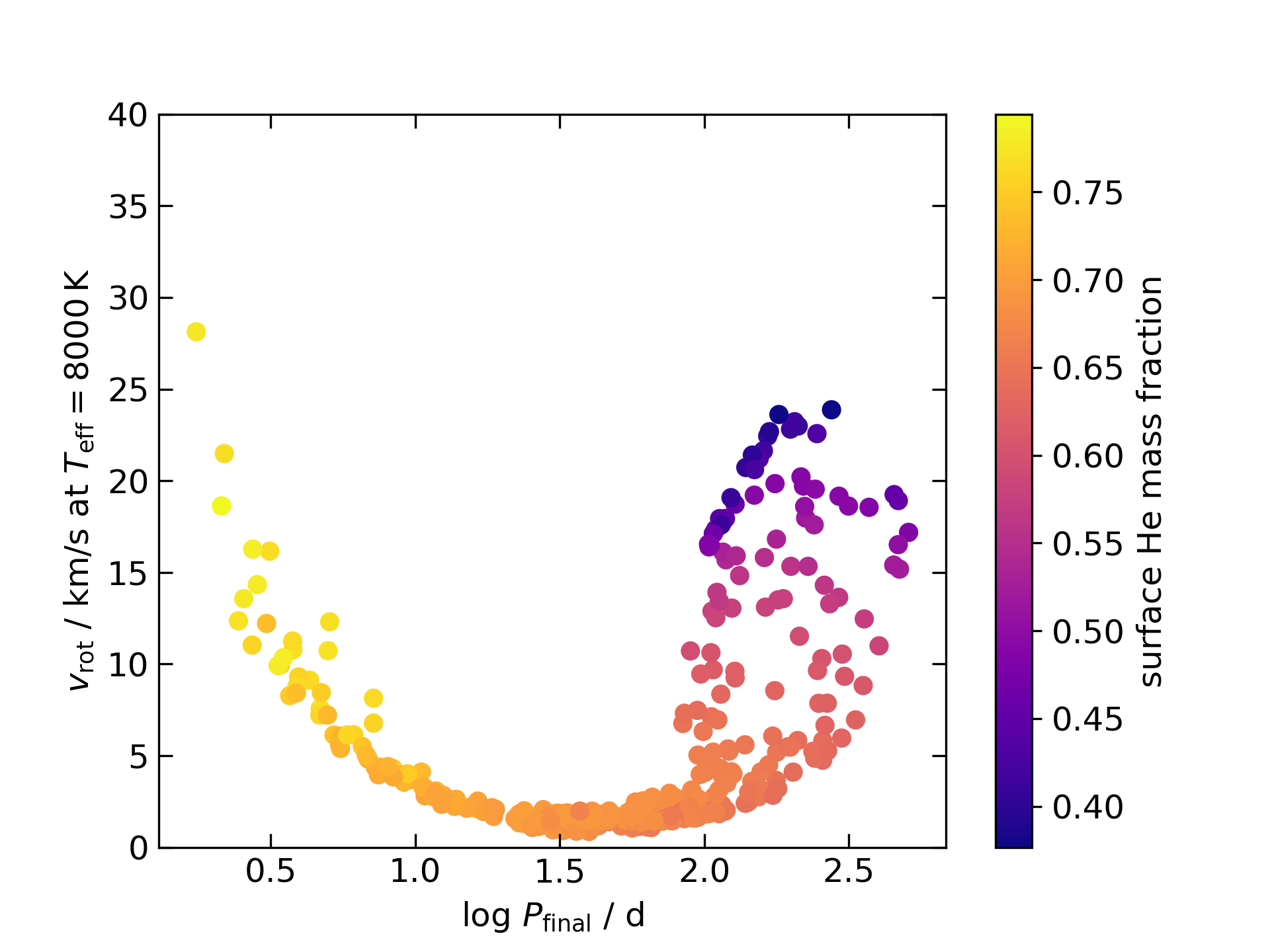}
    \includegraphics[width=0.5\hsize]{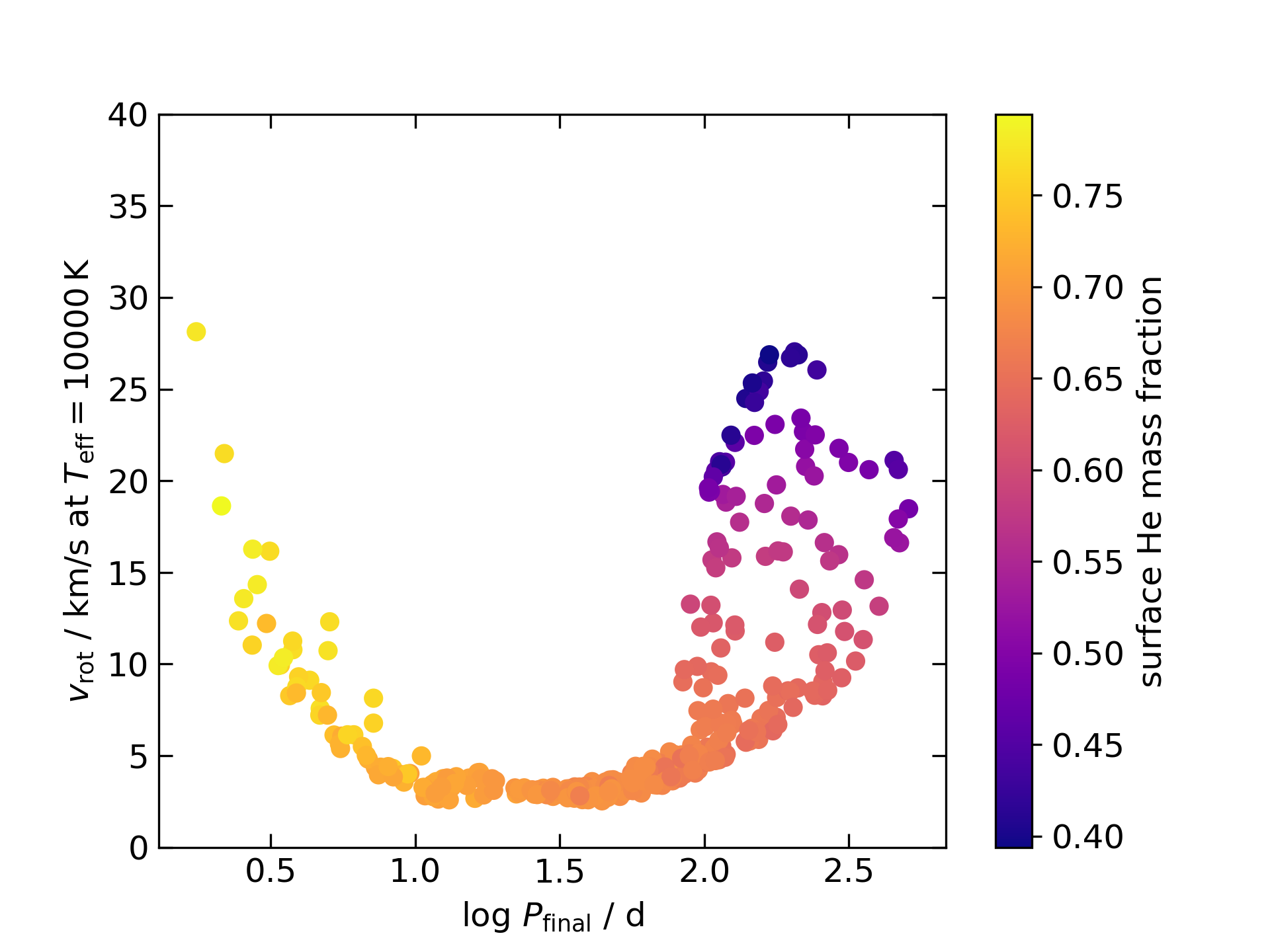}
    \includegraphics[width=0.5\hsize]{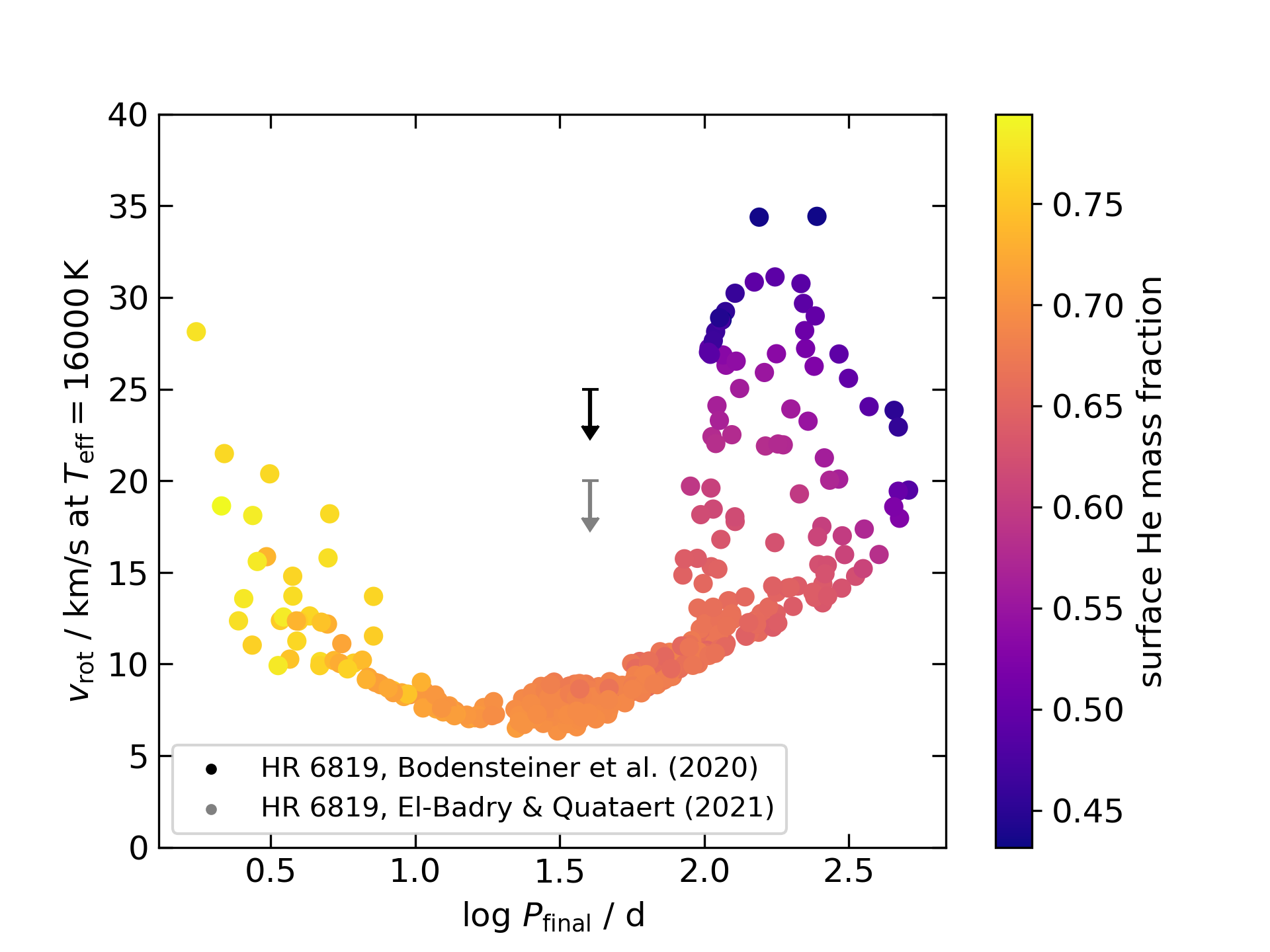}
    \includegraphics[width=0.5\hsize]{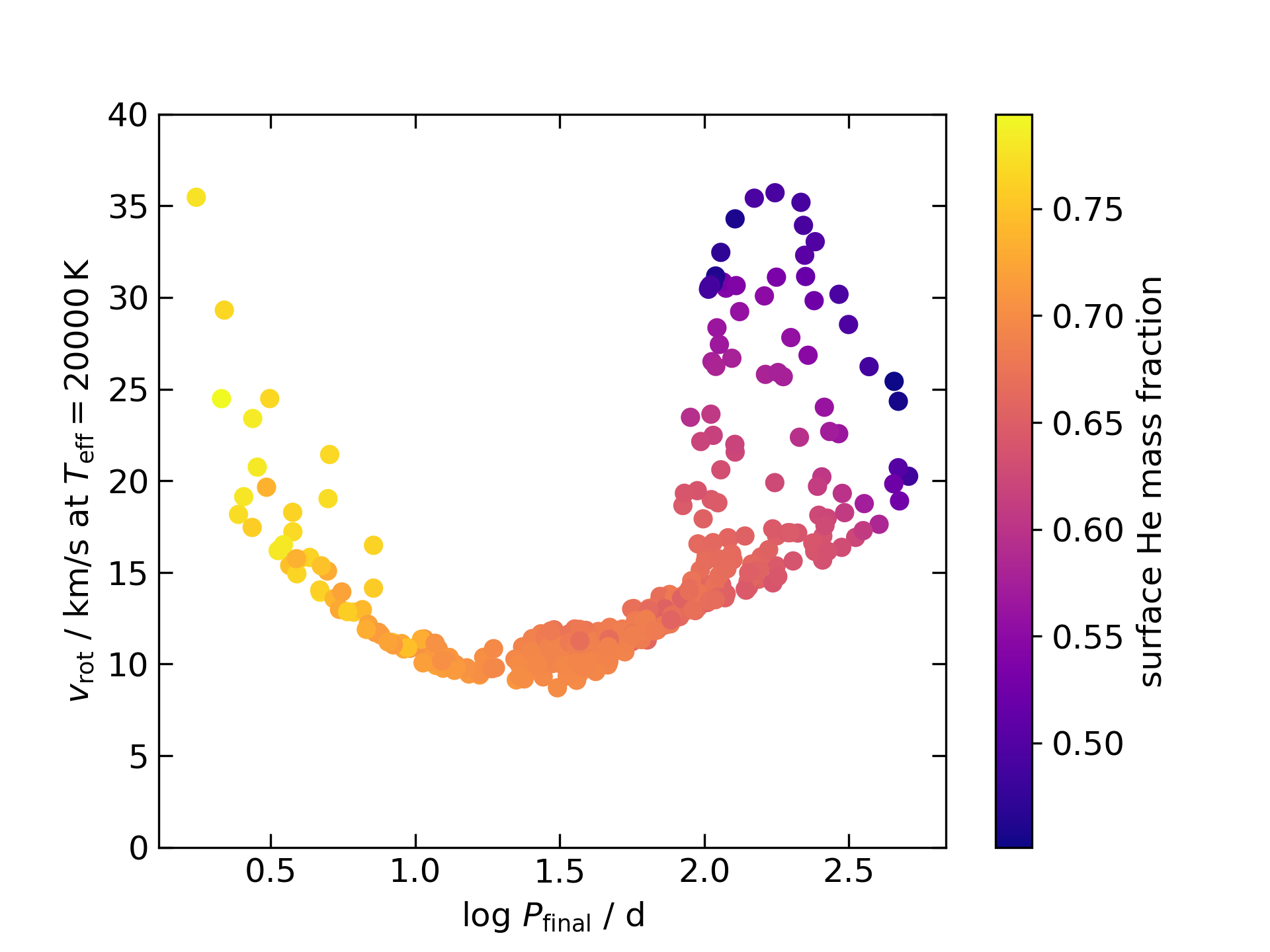}
    \caption{Same as Fig.~\ref{fig:prog}, but with an effective temperature of 8\,000\,K (top left), 10\,000\,K (top right), 16\,000\,K (bottom left), and  20\,000\,K (bottom right). The bottom left panel shows the measurements of \citet{bodensteiner} and \citet{elbadry2} for HR~6819.}
    \label{fig:prog2}
\end{figure*}

%\listofobjects

\end{document}